\documentclass[11pt,fleqn,a4paper]{article}

\usepackage{amsmath,amssymb,amsthm,comment,epsfig,graphics,backref}
\usepackage[mathscr]{eucal}
\usepackage[labelsep=period]{caption}
\usepackage[utf8]{inputenc}
\usepackage[english]{babel}
\usepackage{float}

\usepackage{xcolor}

\usepackage{hyperref}

\newcommand{\p}{{\partial}}

\newtheorem{theorem}{Theorem}

\newtheorem{proposition}[theorem]{Proposition}
\newtheorem*{proposition*}{Proposition}
{\theoremstyle{definition}
\newtheorem{definition}[theorem]{Definition}

\newtheorem{remark}[theorem]{Remark}

}

\newcommand{\todo}[1][\null]{\ensuremath{\clubsuit}}

\newcommand{\noprint}[1]{}

\DeclareMathOperator{\ad}{ad}

\DeclareMathOperator{\V}{Vec_{hol}}
\DeclareMathOperator{\kerR}{ker}

\begin{document}

\noindent
UDC: 512.816:517.92

\vspace{2mm}\par\noindent
{\large M.\,S.~Staryi} \ ORCID: 0009-0002-7442-5813
\par\vspace{1mm}\par\noindent
{\small Institute of Mathematics of the National Academy of Sciences of Ukraine, Kyiv, Ukraine
\par\noindent
E-mail: staryimath@gmail.com}

\par\vspace{2mm}\par\noindent
{\large M.\,O.~Nesterenko} \ ORCID: 0000-0001-6352-2365
\par\vspace{1mm}\par\noindent
{\small Institute of Mathematics of the National Academy of Sciences of Ukraine, Kyiv, Ukraine\\
University ``Kyiv School of Economics'', Kyiv, Ukraine
\par\noindent
E-mail: maryna.nesterenko@gmail.com}

\par\vspace{5mm}\par\noindent
{\LARGE\bf Contractions of realizations and differential invariants
of the algebra $\mathfrak{sl}(2,\mathbb{C})$}

\par\vspace{2mm}\par\noindent

\par\vspace{4mm}\par\noindent
{\small\it
In this paper, we formalize a method for constructing limit transitions
between local realizations of Lie algebras based on contractions of the
corresponding algebras. In the complex holomorphic category, contractions
of faithful realizations of the special linear algebra $\mathfrak{sl}(2,\mathbb{C})$
to the complex three-dimensional Poincar\'e algebra and the Weyl--Heisenberg
algebra are explicitly constructed along the chain
$\mathfrak{sl}(2,\mathbb{C}) \to \mathfrak{g}_{3.4}^{-1} \to \mathfrak{g}_{3.1}$.

For the contraction of a generating realization of the algebra
$\mathfrak{sl}(2,\mathbb{C})$ to the algebra $\mathfrak{g}_{3.4}^{-1}$,
differential invariants are constructed and a limit system of Newtonian type
is found.}

\par\vspace{2mm}\par\noindent

\par\vspace{6mm}

The concept of a contraction, or limit transition between two
mathematical structures of the same type,
was first introduced by I.~Segal in 1951~\cite{Segal1951}
as a purely mathematical construction; a survey of further developments
in the study of contractions can be found in~\cite{NP2006, Popovych2009}.
Contractions gained widespread application following the work of
In\"on\"u and Wigner~\cite{Inonu1953}, who formalized the notion of a
limit transition between different physical theories by introducing
contractions of Lie algebras associated with subalgebras
(the so-called \emph{In\"on\"u--Wigner contractions}).
One of the best-known examples of a contraction is the relation between
relativistic and classical mechanics, which are invariant under the Poincar\'e
and Galilei groups, respectively; the corresponding limit transitions relate,
in particular, the Klein--Gordon equation and the Schr\"odinger
equation~\cite{Inonu1953}.
Another classical example~\cite{Segal1951} is the contraction of the
three-dimensional special linear algebra $\mathfrak{sl}(2)$ to the
Weyl--Heisenberg algebra $\mathfrak{h}_3=\mathfrak{g}_{3.1}$;
similar limit transitions are used to establish connections between
quantum-mechanical models.

Thus, by studying contractions, one can find limit transitions between
differential equations invariant under the corresponding Lie groups or Lie algebras.
Examples of finding limit transitions between realizations of Lie algebras and
invariant equations have already appeared in the
literature~\cite{Bluman, Ivanova2007, NPV2016, NesterenkoPosta2020, NesterenkoPostaStaryi2025},
where a number of concrete examples and certain theoretical aspects were considered.
In the present article, we focus on formalizing the procedure for constructing
parameterized realizations and the invariant systems associated with them.

The main goal of this work is to formalize a practical approach that allows
one to relate invariant models by means of known limit transitions between
their symmetry algebras, and to test this algorithm on the example of
three-dimensional Lie algebras.
The theoretical treatment is accompanied by examples: proper nontrivial
contractions of the algebra $\mathfrak{sl}(2,\mathbb{C})$ are investigated,
and for selected contractions, families of parameterized realizations and
their limits are constructed.
Lie algebras are considered over the complex field, and their local realizations
are realized by holomorphic vector fields on complex manifolds. Equivalence of
realizations is understood locally, using biholomorphic coordinate transformations.
At the end of the paper, we also briefly outline the conceptual framework
required to describe invariant equations via differential invariants.
For one parameterized realization of the algebra $\mathfrak{sl}(2,\mathbb{C})$,
we consider an example of constructing differential invariants, invariant systems,
and a limit transition between systems of two second-order ordinary differential
equations with one independent and two dependent variables of the form
\begin{gather}\label{gen_sys}
\left\{
  \begin{gathered}
    x_{tt} = F(t,x,y,x_t,y_t), \\
    y_{tt} = G(t,x,y,x_t,y_t).
  \end{gathered}
\right.
\end{gather}
Such equations belong to systems of Newtonian type, which are actively used
to model the dynamics of classical and quantum systems in external force fields.
In the real case, system~\eqref{gen_sys} can be viewed as a generalized form
of Newton's second law for a system with two degrees of freedom.
At the same time, the functions $F$ and $G$ play the role of components of an
external force that may explicitly depend on time and velocities.
Such forces are characteristic of problems with dissipation or of describing
the motion of a charged particle under the Lorentz force in an electromagnetic field.
Among important examples of such dissipative systems are the van der Pol equation
(modeling self-oscillations in electrical circuits) and its generalization---the
classical Li\'enard equation for systems with nonlinear friction (for more details
on the symmetry properties of this equation, see~\cite{ML1990,Tiwari2013}).
Another example of a nonlinear model is the Duffing oscillator, which is used,
in particular, to describe nanoresonators~\cite{LC2007}.
These physical examples are formulated primarily in the real category; comparing
them with the results of the present work requires an appropriate complexification.

In general, integrating nonlinear systems of Newtonian type is a difficult problem.
However, for certain classes of functions $F$ and $G$, it can be substantially
simplified or reduced to quadratures using a group-theoretic approach.
A fundamental problem in this context is the group classification of equations
of this type~\cite{Ovsiannikov1982, Olver1993, Fushchich1993}, i.e., identifying
systems that admit nontrivial Lie algebras of point symmetries.

An important subclass of such equations consists of nonlinear systems of
Ermakov type, which arise in many problems of applied physics, in particular
in describing the evolution of charged-beam parameters in nonstationary
electromagnetic fields~\cite{FTJNR2013}. In~\cite{HG2004}, it was shown that
the radial equation of motion of an ion in a Paul radio-frequency trap reduces
to an Ermakov equation admitting the point symmetry group $\mathrm{SL}(2)$.
The appeal of Ermakov systems stems from the existence of a special first
integral---a quantity constant along solutions---traditionally called the
Ermakov invariant.
In this sense, the term ``invariant'' should not be automatically identified
with a differential invariant of a group action.
Such systems also admit the construction of solutions via a nonlinear
superposition principle~\cite{RR1980,ARRO1990}.
This integral is also related to the Lewis invariant for an oscillator with
time-dependent frequency~\cite{Lewis1967}.

The paper is organized as follows.
Section~1 provides the necessary theoretical background.
Section~2 is devoted to a detailed study of the chain of algebra contractions
\[
\mathfrak{sl}(2,\mathbb{C})
\to \mathfrak{g}_{3.4}^{-1}
\to \mathfrak{g}_{3.1}.
\]
In Section~3, parameterized families of local realizations are constructed
by an algebraic method, and explicit transformations relating these realizations
to canonical representatives of the corresponding algebras are found.
In Section~4, for a single example, we compute a basis of first- and
second-order differential invariants, construct invariant systems of
differential equations, and investigate the limit transition.
The resulting invariant system is also compared with the system obtained
in~\cite{Haponova2006}.

\section{Lie algebras, their contractions and realizations}
In this section, to make the exposition self-contained, we review the necessary
theoretical background on Lie algebras, their contractions, and their realizations,
which is well known and can be found in~\cite{NP2006,PBNL2003, Wells2008, PopovychPhD}
and the references therein.

Consider an $n$-dimensional complex vector space $V$ and a \emph{Lie algebra}
$\mathfrak{g}$ on $V$, which is a pair $(V,[\cdot,\cdot])$, where $[\cdot,\cdot]$
is a bilinear skew-symmetric operation satisfying the Jacobi identity.
This operation is also called the Lie bracket or commutator.

Fix a basis $e_1, \dots, e_n$ of the vector space $V$; then the Lie algebra
$\mathfrak{g}=(V,[\cdot,\cdot])$ is uniquely determined by the tensor of
structure constants of type $(1,2)$, namely
$c_{ij}^k\in \mathbb{C}$, $i, j, k\in \{1,\dots, n\}$,
which defines the commutators of the basis elements:
$[e_i,\,e_j]=\sum\limits_{k=1}^n c_{ij}^k e_k.$

The group $\mathrm{GL}(V)$ of nonsingular complex linear transformations acts on
the vector space $V$, partitioning the variety of all complex Lie algebras of a
given dimension into isomorphism classes. Under each transformation $U\in\mathrm{GL}(V)$
with matrix $(U_{j}^i)_{i,j=1}^n$, the tensor of structure constants transforms
according to the rule
$\tilde{c}_{i'j'}^{k'}=\sum\limits_{i,j,k=1}^n \,(U^{-1})_k^{k'}U_{i'}^{i} U_{j'}^{j}c_{ij}^k$,
where $i', j', k'\in \{1,\dots, n\}$.

The study of properties and applications of Lie algebras is closely tied to
the classification of all inequivalent Lie algebras of a given dimension over
a given field. In this paper, we do not dwell on a detailed analysis or survey
of existing classifications, but we note that we adhere to the numbering consistent
with the classification of three-dimensional real Lie algebras used in~\cite{PBNL2003}.
This choice of numbering is motivated by the need to apply our previous results
and compare the systems and invariants obtained in the limit cases.

Below is the complete list of inequivalent complex three-dimensional Lie algebras
together with their nonzero commutation relations:
\begin{equation}\label{classification3d}
\begin{array}{ll}
3\mathfrak{g}_1\colon &  \text{(Abelian algebra)};\\
\mathfrak{g}_{2.1}\oplus\mathfrak{g}_1\colon & [e_1, e_2] = e_1; \\
\mathfrak{g}_{3.1}\colon & [e_2, e_3] = e_1\ \text{(Heisenberg algebra and reduced classical Galilei algebra)}; \\
\mathfrak{g}_{3.2}\colon & [e_1, e_3] = e_1,\ [e_2, e_3] = e_1 + e_2; \\
\mathfrak{g}_{3.3}\colon & [e_1, e_3] = e_1,\ [e_2, e_3] = e_2; \\
\mathfrak{g}_{3.4}^a\colon & [e_1, e_3] = e_1,\ [e_2, e_3] = a e_2,\ a\in\mathbb{C}\setminus\{0,1,-1\},\ a\sim a^{-1}; \\
\mathfrak{g}_{3.4}^{-1}\colon & [e_1, e_3] = e_1,\ [e_2, e_3] = -e_2\ \text{(three-dimensional Poincar\'e algebra)}; \\
\mathfrak{sl}(2,\mathbb{C})\colon & [e_1, e_2] = e_1,\ [e_1, e_3] = 2e_2,\ [e_2, e_3] = e_3\ \text{(special linear algebra)}. \\
\end{array}
\end{equation}

Over the field of real numbers, the classification differs from the complex one,
since non-isomorphic real algebras may have isomorphic complexifications.
In particular, the complexification of the rotation group algebra $\mathfrak{so}(3)$
is isomorphic to $\mathfrak{sl}(2,\mathbb{C})$.
The complexifications of the algebras in the family $\mathfrak{g}_{3.5}^b$, under
an appropriate matching of parameters, belong to the family $\mathfrak{g}_{3.4}^{a}$.

The next natural question is the study of relations between different isomorphism
classes of Lie algebras of a given dimension. One of the fundamental mechanisms
describing transitions from one algebra to another is a Lie algebra contraction---a
special limit of a parametric deformation of its structure constants under which
the initial algebra degenerates into a new one, typically less complex and
non-isomorphic to it.
Studying contractions is important for several reasons: they establish a hierarchy
among classes of Lie algebras, have direct applications in mathematical physics,
and allow one to build a complete picture of orbit closures in the variety of
algebras of a given dimension, which is an important problem of algebraic geometry
in its own right.

Let $U\colon(0,1]\to\mathrm{GL}(V)$ be a continuous matrix-valued function defining
a parameterized family of nonsingular linear transformations
$U(\varepsilon)\in\mathrm{GL}(V)$, $\varepsilon\in(0,1]$.
Consider the one-parameter family of Lie algebras
$\mathfrak{g}_\varepsilon=(V,[\cdot,\cdot]_\varepsilon)$ with commutation relations
\begin{equation}\label{eq:eps_bracket}
  [v,w]_\varepsilon := U(\varepsilon)^{-1}\bigl[U(\varepsilon)v,\,U(\varepsilon)w\bigr],
  \qquad v,w\in\mathfrak g,\ \ \varepsilon\in(0,1].
\end{equation}

\begin{definition}
If for all $v,w\in V$ the limit
$[v,w]_0:=\lim_{\varepsilon\to 0+} [v,w]_\varepsilon$
exists, then the Lie bracket $[\cdot,\cdot]_0$ is well defined, and the corresponding
Lie algebra $\mathfrak{g}_0=(V,[\cdot,\cdot]_0)$ is called a
\emph{contraction} of the algebra $\mathfrak{g}$, while the parameter $\varepsilon$
is called the \emph{contraction parameter}.
By a contraction we also mean the limit transition between the algebras
$\mathfrak{g}\to \mathfrak{g}_0$.

If the algebra $\mathfrak{g}_0$ is isomorphic to an Abelian algebra, the contraction
$\mathfrak{g}\to \mathfrak{g}_0$ is called \emph{trivial};
if $\mathfrak{g}_0$ is isomorphic to the original algebra $\mathfrak{g}$,
the contraction is called \emph{improper}.
\end{definition}

In a fixed basis of the algebra $\mathfrak{g}$, the parameterized transformation
$U(\varepsilon)$ can be represented by a matrix called the \emph{contraction matrix}.

In the basis $e_1,\dots,e_n$, it follows immediately from~\eqref{eq:eps_bracket} that
\begin{equation}\label{tens_str_eps}
  \tilde{c}_{i'j'}^{k'}(\varepsilon)=\sum\limits_{i,j,k} U(\varepsilon)_{i'}^i
  U(\varepsilon)_{j'}^j  (U(\varepsilon)^{-1})_k^{k'} c_{ij}^k.
\end{equation}

For each $\varepsilon>0$, formula~\eqref{tens_str_eps} coincides with the
transformation rule for structure constants under the change to the basis
$U(\varepsilon)e_1,\dots,U(\varepsilon)e_n$ of the initial algebra.
Therefore, the algebra $\mathfrak g_\varepsilon$ is isomorphic to $\mathfrak g$.
At the same time, the limit algebra $\mathfrak g_0$ need not be isomorphic
to $\mathfrak g$: the limit transition of structure constants does not necessarily
correspond to a nonsingular change of basis.

In this case, the structure constants $\hat{c}_{ij}^k$ of the resulting algebra
$\mathfrak{g}_0$ are determined by the limit
$\hat{c}_{ij}^k=\lim\limits_{\varepsilon\to 0+}\tilde{c}_{ij}^k(\varepsilon).$

From the viewpoint of algebraic geometry, a Lie algebra contraction is a limit
point of its orbit in the Zariski topology.
Moreover, for orbits of the $\mathrm{GL}(n,\mathbb{C})$-action on the variety
$\mathcal{L}_n$ of complex $n$-dimensional Lie algebras, the Zariski closure
coincides with the metric closure.
The distinction between these notions lies in the fact that a contraction is a
specific mechanism of limit transition, whereas the orbit closure provides the
complete topological-geometric picture of all possible degenerations.
Rigid algebras, whose orbits are open in the Zariski topology, form the skeleton
of the variety $\mathcal{L}_n$, while chains of contractions define a partial
order relation stratifying the space by complexity.

In general, finding a contraction matrix for representatives of fixed Lie algebras
is a technically challenging problem, although D.~Popovych~\cite{PopovychPhD}
proposed a constructive approach to solving it.

In certain cases, including contractions of the algebra $\mathfrak{sl}(2,\mathbb{C})$,
one can restrict attention to matrices diagonal in the contraction parameter---the
so-called In\"on\"u--Wigner contractions.

\begin{definition}
Suppose a contraction $\mathfrak{g}\to \mathfrak{g}_0$ is defined by a matrix
$U(\varepsilon)$ that can be represented in the form
$U(\varepsilon)=I_0W(\varepsilon)I$, where $I$ and $I_0$ are constant
nonsingular matrices and
$W(\varepsilon)=\mathrm{diag}(\varepsilon^{\alpha_1},\dots,\varepsilon^{\alpha_n})$.
The numbers $\alpha_i\in \mathbb{Z}$, $i=1,\dots, n$, are called the \emph{signature}
of the contraction, and the contraction is called a \emph{generalized In\"on\"u--Wigner
contraction}, or GIW contraction.

In the case $\alpha_i\in \{0, 1\}$, $i=1,\dots, n$, the contraction is called a
\emph{simple In\"on\"u--Wigner contraction}, or IW contraction.
The matrices $I$ and $I_0$ do not affect the signature; see Remark~\ref{rem:AB}.
\end{definition}

\begin{remark}\label{rem:AB}
Let $S,T\in\mathrm{GL}(V)$ be constant and $U(\varepsilon)=S\widetilde U(\varepsilon)T$.
Denote by $[\cdot,\cdot]^{(S)}$ the bracket $[v,w]^{(S)}=S^{-1}[Sv,Sw]$.
Then
$[v,w]_{\varepsilon}^{(U)} = T^{-1}\left([Tv,Tw]_{\varepsilon}^{(\widetilde U;\,[\cdot,\cdot]^{(S)})}\right)$,
where the superscript on the right indicates that in constructing the parameterized
bracket via $\widetilde U(\varepsilon)$, the initial bracket is $[\cdot,\cdot]^{(S)}$.
Hence, the right factor $T$ defines a constant change of basis for the parameterized
brackets and, when the limit exists, for the limit brackets.
The left factor $S$, in general, alters the initial bracket in the computation and
cannot be discarded unless $S$ is an automorphism of it.
\end{remark}

For a generalized In\"on\"u--Wigner contraction, the weights $\alpha_i$ define a
decomposition of the vector space into weight subspaces.
In a basis in which the contraction matrix is diagonal, the corresponding structure
constants have the form
\[
\tilde c_{ij}^{\,k}(\varepsilon)
=
\varepsilon^{\alpha_i+\alpha_j-\alpha_k}c_{ij}^{\,k}.
\]
The limit exists if and only if $\alpha_i+\alpha_j-\alpha_k\geqslant 0$ for all
$c_{ij}^{\,k}\ne 0$. In the limit bracket, precisely those terms remain for which
$\alpha_i+\alpha_j=\alpha_k$.
The weight decomposition defines a filtration of the Lie algebra, and the limit
algebra is the corresponding associated graded algebra.

For a simple In\"on\"u--Wigner contraction, we have the decomposition
$V=V_0\oplus V_1$, where the indices correspond to weights $0$ and $1$.
Here, $V_0$ is a subalgebra of the limit algebra,
$[V_0,V_1]_0\subseteq V_1$, and $[V_1,V_1]_0=0$.

The primary tools that allow one to rule out pairs of algebras between which no
contraction is possible are necessary criteria based on the semicontinuity of
algebraic invariants (dimensions of derived series, ranks of adjoint operators,
dimension of the derivation algebra, etc.), which exhibit monotonic behavior under
degeneration.
Precisely these approaches will be explicitly implemented in Section~\ref{Sec_contractions}
to find all possible contractions of the algebra $\mathfrak{sl}(2, \mathbb{C})$
and to construct the corresponding matrices explicitly.

We also state one more proposition needed to justify contractions of realizations
constructed by the algebraic method from a subalgebra.
\begin{proposition}\label{TheoremOnContractionsOfSubalgs}
Let a Lie algebra $\mathfrak g_0$ be a contraction of a Lie algebra $\mathfrak g$,
and suppose $\mathfrak g$ contains a subalgebra $\mathfrak h$.
Then $\mathfrak g_0$ contains a subalgebra $\mathfrak h_0$ of the same dimension,
$\dim\mathfrak h_0=\dim\mathfrak h$.
\end{proposition}

The concepts presented above describe a Lie algebra as an abstract object---through
the commutation relations among its basis elements---and relate different algebras
to one another.
However, for applications in the theory of differential equations, this abstract
picture is not sufficient. Since our ultimate goal is for a Lie algebra to act
as the symmetry algebra of concrete equations, its elements must be realized as
infinitesimal generators of point transformations on the space of independent
and dependent variables.
Such a representation expresses elements of the Lie algebra as vector fields
and is called a realization of the Lie algebra.
It is precisely the realization that serves as a bridge between an abstract
symmetry algebra and differential equations admitting that symmetry.
Once the explicit form of the vector fields is found, one can construct invariants,
differential invariants, and the corresponding invariant equations, thereby solving
our main problem---the construction of equations with a prescribed symmetry algebra.

Consider a simply connected $m$-dimensional complex manifold $M$, defined as a
Hausdorff topological space with a countable base of topology and an atlas with
holomorphic transition functions between charts.

Let $(z_1, \dots, z_m)$, where $z_j = x_j + i y_j$, $j=1,\dots,m$, denote local
complex coordinates on $M$. Then the complexified tangent space $T_p M \otimes \mathbb{C}$
at a point $p \in M$ has two natural bases:
the real basis $\{ \frac{\partial}{\partial x_1}, \frac{\partial}{\partial y_1}, \dots,
\frac{\partial}{\partial x_m}, \frac{\partial}{\partial y_m} \}$
and the complex Wirtinger basis\linebreak
$\tfrac{\partial}{\partial z_j} = \frac{1}{2} \left( \tfrac{\partial}{\partial x_j} - i \tfrac{\partial}{\partial y_j} \right)$,
$\tfrac{\partial}{\partial \bar{z}_j} = \frac{1}{2} \left( \tfrac{\partial}{\partial x_j} + i \tfrac{\partial}{\partial y_j}\right)$.

Locally, any complex vector field $\Xi$ can be written with smooth complex-valued
coefficients ${\xi}_j$ and $\check{\xi}_j$ in the form
\mbox{$\Xi = \sum_{j=1}^{m} \left( \xi_j(z, \bar{z}) \frac{\partial}{\partial z_j} +
\check{\xi}_j(z, \bar{z}) \frac{\partial}{\partial \bar{z}_j} \right)$}.

Vector fields are said to be equivalent if they are related by a change to another
system of holomorphic coordinates (we consider only holomorphic vector fields, i.e.,
fields of type $(1,0)$ whose antiholomorphic part vanishes and whose coefficients
in the holomorphic basis are holomorphic functions).
In an arbitrary holomorphic coordinate chart, a holomorphic vector field can be
written as
\begin{gather*}\label{eq_vect_field}
\Xi = \sum_{j=1}^{m} \xi_j(z) \frac{\partial}{\partial z_j},\quad
\text{where}\ \xi_j(z)=\xi_j(z_1, \dots, z_m)\ \text{are holomorphic functions.}
\end{gather*}

In another system of holomorphic coordinates $w = (w_1, \dots, w_m)$, the new
coefficients are obtained via the chain rule
by substituting into the expression for $\Xi$:
$\tilde{\xi}_k(w) = \sum\limits_{j=1}^{m} \xi_j(z) \frac{\partial w_k}{\partial z_j}$.

Let $M_p \subset M$ be an open neighborhood of the point $p$, and let $\V(M_p)$
denote the space of holomorphic vector fields on $M_p$.
The Lie bracket of two holomorphic vector fields is again a holomorphic vector field;
hence $\V(M_p)$ is a Lie algebra over $\mathbb{C}$.
Consider homomorphisms from the algebra $\mathfrak{g}$ into $\V(M_p)$, i.e.,
assign to a basis of $\mathfrak{g}$ a collection of local vector fields
$\Xi_i = \xi_i^a(z) \partial_{z_a}$ whose commutators $[\Xi_i,\Xi_j]$ reproduce
the given structure constants of the algebra.

\begin{definition}
A \emph{realization} of a complex Lie algebra $\mathfrak{g}$ in a neighborhood of
a point $p \in M$ is a homomorphism $R\colon\mathfrak{g}\to\V(M_p)$.

A realization is called \emph{faithful} if $R$ is injective
(i.e., $\kerR R = \{0\}$), and \emph{degenerate} if the vector fields
$R(e_1),\dots,R(e_n)$ are linearly dependent over $\mathbb{C}$ with \emph{constant}
coefficients.
\end{definition}

\begin{remark}\label{rem:exact_vs_rank}
Linear dependence over the ring of holomorphic functions $\mathcal O(M)$ concerns
the rank rather than faithfulness. The realization
$\partial_1,\ x_1\partial_1,\ x_1^2\partial_1$ of the algebra $\mathfrak{sl}(2,\mathbb C)$
on $\mathbb C$ is faithful ($\kerR R=\{0\}$), even though its fields are
functionally dependent: $x_1^2R(e_1)-2x_1R(e_2)+R(e_3)=0$.
\end{remark}

Given two local realizations $R_1$ and $R_2$ in neighborhoods of points $p_1$ and $p_2$
(neighborhoods $M_1$ and $M_2$, respectively),
$R_1\colon\mathfrak{g}\longrightarrow\V(M_1)$ and
$R_2\colon \mathfrak{g}\longrightarrow\V(M_2)$,
establishing their equivalence requires accounting for both geometric holomorphic
transformations of the realization domains and complex-linear automorphisms of
the algebra $\mathfrak{g}$.
In the real case, realizations are compared via diffeomorphisms; in the complex
case, one must impose the stricter requirement of a biholomorphic transformation
between coordinates.
Denote the automorphism group of the algebra $\mathfrak{g}$ by
$\mathrm{Aut}(\mathfrak{g})=\left\{\psi\in\mathrm{GL}(\mathfrak{g})\middle|\psi([x,y])=[\psi(x),\psi(y)]\right\}$.

\begin{definition}
Realizations $R_1\colon\mathfrak{g}\to\V(M_1)$ and $R_2\colon\mathfrak{g}\to\V(M_2)$
are called \emph{weakly equivalent} if there exist an automorphism
$\psi\in \mathrm{Aut}(\mathfrak{g})$ and a biholomorphism
$\Phi\colon M_1\to M_2$ such that
$R_2(\psi(v))=\Phi_* R_1(v)$ for all $v\in \mathfrak{g}$.
If $\psi$ is the identity map, the realizations are mapped into one another solely
by a local biholomorphism. In this special case, the realizations are called
\emph{strongly} equivalent.
\end{definition}
Other types of equivalence can also be defined.

An important difference between the real~\cite{PBNL2003} and complex cases is
the local nature of a complex realization. However, since $M$ is a connected complex
manifold, a holomorphic vector field that vanishes on a nonempty open subset
vanishes on the entire connected component of its domain.
This yields the following consequences:
\begin{itemize}
\item if two holomorphic realizations coincide on an open subset of the intersection
of their domains, they coincide on the entire corresponding connected component;
\item if a local holomorphic continuation of a realization exists, it is unique.
\end{itemize}
At the same time, a local realization does not always extend automatically to the
entire manifold: global continuation requires taking into account the topology of
the manifold, monodromy, and possible analytic obstructions.

\begin{definition}
The \emph{rank of a realization at a point $p$} is defined as
$\operatorname{rank}R_p=\dim_{\mathbb C}\operatorname{span}_{\mathbb C}\{R(v)|_p\}_{v\in\mathfrak g}$,
and $\operatorname{rank}R=\max_{p}\operatorname{rank}R_p$.

A realization $R$ is called \emph{transitive} if $\operatorname{rank}R_p=m$ for
\emph{every} point of its domain, and \emph{regular} if $\operatorname{rank}R_p$
is locally constant; the regular rank may be smaller than $m$.
\end{definition}

Note that the condition $\max_p\operatorname{rank}R_p=m$ yields transitivity only on
the open subset $\{p:\operatorname{rank}R_p=m\}$, which may be proper. For example,
the field $x\partial_x$ on $\mathbb C$ has maximal rank $1$, while the point $0$
is fixed.
The rank of a realization is conveniently computed by forming the matrix of coefficients
of the vector fields.

For a subset $\mathfrak h\subseteq\mathfrak g$, define
\[
\operatorname{rank}R(\mathfrak h)_p
=
\dim_{\mathbb C}\operatorname{span}_{\mathbb C}
\{R(v)|_p\}_{v\in\mathfrak h}.
\]
If $\mathfrak h$ is invariant under all automorphisms of the algebra $\mathfrak g$,
then its rank at corresponding points is preserved under weak equivalence.
Consequently, a discrepancy between these ranks at corresponding points provides
a criterion for inequivalence. For an arbitrary subset, one must also take into
account its image under the action of an automorphism.

The next task toward constructing systems of Newtonian type is the explicit
construction of representatives of equivalence classes.
The problem of classifying realizations splits into two interrelated parts:
constructing a system of vector fields realizing a given abstract Lie algebra,
and isolating all inequivalent cases. This topic has been widely addressed in the
literature, yet no complete algorithm for constructing all realizations currently
exists. We consider two approaches to constructing realizations, the choice between
which is dictated by the specifics of a given problem.

The direct method~\cite{PBNL2003} starts from the classification of abstract Lie
algebras and uses the technique of megaideals to reduce the problem.
The direct construction of realizations reduces to solving systems of partial
differential equations for the coefficient functions of the vector fields, followed
by classifying the resulting realizations up to weak equivalence.
This stage requires computing automorphism groups of the algebra, finding its
ideals, and analyzing admissible nonsingular changes of variables.

The algebraic method~\cite{MMS2015}, proposed by Magazev, Mikheyev, and Shirokov,
proceeds from a fundamentally different idea: a realization is constructed as a
set of left-invariant vector fields on the Lie group itself using solely the
structure constants of the algebra.
In canonical coordinates of the second kind, the components of left-invariant
1-forms $\omega$ are computed via the matrices of the adjoint representation,
after which the corresponding vector fields are found by inverting the matrix $\omega$.
Thus, the problem reduces to purely linear-algebraic operations---computing matrix
exponentials and matrix inverses. The method produces transitive realizations:
simply transitive on $G$ (the case $\mathfrak h=0$) and transitive on homogeneous
spaces $G/H$.

Note that all subsequent constructions are carried out in the complex holomorphic
category, where $M$ is a complex manifold with $\dim_{\mathbb C}M=m$, and the
fields of a realization are holomorphic sections of $T^{1,0}M$. The complexification
of the real tangent space $T_pM\otimes_{\mathbb R}\mathbb C$ (of dimension $2m$)
is not used here.
All constructions are local, so a coordinate neighborhood suffices; what is
essential is not locality itself, but the identity theorem for holomorphic coefficients.

By the classical theory of Sophus Lie, vector fields realizing a Lie algebra
$\mathfrak{g}$ can be identified with the infinitesimal generators of a local
Lie group $G$ of transformations of the manifold $M$.
In~\cite{MMS2015}, it is shown that any transitive local realization of a Lie
algebra $\mathfrak{g}$ by vector fields on an $m$-dimensional manifold $M$ is
determined by a subalgebra of codimension $m$,
$\mathfrak{h} = \{v \in \mathfrak{g} \mid R(v)|_p = 0\}$,
which is the stabilizer of the point $p \in M$.
That is, there is a one-to-one correspondence between classes of locally
equivalent transitive realizations of $\mathfrak{g}$ and conjugacy classes of
subalgebras $\mathfrak{h}$.

From a geometric standpoint, this follows from the fact that any transitive action
of a Lie group $G$ is locally equivalent to the action on the space of left cosets
$G/H$ by the subgroup $H$ corresponding to the subalgebra $\mathfrak{h}$.

Let $R$ be a transitive realization on $M$ and let
$\mathfrak h=\{v\in\mathfrak g\mid R(v)|_p=0\}$ be the stabilizer of a point $p$;
then $\operatorname{codim}\mathfrak h=m$.
According to~\cite{MMS2015}, $\kerR R=\operatorname{core}_{\mathfrak g}(\mathfrak h)$,
where $\operatorname{core}_{\mathfrak g}(\mathfrak h)$ is the largest ideal of
$\mathfrak g$ contained in $\mathfrak h$.
In particular, the realization $R$ is faithful if and only if $\mathfrak h$
contains no nonzero ideals of $\mathfrak g$.
The classification of transitive realizations reduces to the classification of
subalgebras of codimension $m$, where
$\mathfrak h'=\mathfrak h \iff R'\ \text{is strongly equivalent to}\ R$,
$\mathfrak h'\in\operatorname{Ad}_G\mathfrak h \iff R'$ is equivalent to $R$ up to $\operatorname{Inn}(\mathfrak g)$, and
$\mathfrak h'\in\operatorname{Aut}(\mathfrak g)\mathfrak h \iff R'$ is weakly equivalent to $R$.

To construct a realization explicitly via invariant vector fields on a Lie group,
canonical coordinates of the second kind $(x_1, \dots, x_n)$ are used for group
elements $G\ni g = \exp(x_1 e_1)\dots\exp(x_n e_n)$, followed by projection onto
the quotient space.

Let $\mathfrak{h} \subset \mathfrak{g}$ be a subalgebra of codimension $m$.
Choose a basis $(e_1, \dots, e_n)$ of the algebra $\mathfrak{g}$ such that the
vectors $(e_{m+1}, \dots, e_n)$ form a basis of the subalgebra $\mathfrak{h}$.
The component matrix of the Maurer--Cartan 1-form $\omega(x)$ is computed via
the adjoint representation:
\begin{equation}\label{1-form}
  \omega(x)_{ij} = \left( \exp(-x_n \ad_{e_n}) \exp(-x_{n-1} \ad_{e_{n-1}}) \dots \exp(-x_{j+1} \ad_{e_{j+1}}) \right)_{ij},
\end{equation}
where the indices $i$ and $j$ correspond to the component number and the form
number, respectively.
Then the vector fields of the realization in local coordinates $(x_1, \dots, x_m)$
on $G/H$ are obtained as projections of invariant vector fields:
$R(e_j) = \sum\limits_{i=1}^{m} (\omega^{-1}(x_1, \dots, x_m, 0, \dots, 0))_{ij} \frac{\partial}{\partial x_i}.$

This method can be automated in computer algebra systems (Maple was used in the
present work) and also generalized to intransitive cases.
To verify the constructed intransitive realizations, inequivalent representatives
obtained by the direct method~\cite{PBNL2003} were used.
In Section~\ref{Sec_param_realiz}, this algebraic method will be modified and
applied to construct parameterized realizations that yield faithful realizations
after taking the limit in the contraction parameter.


\section{Contractions of the algebra $\mathfrak{sl}(2, \mathbb{C})$}\label{Sec_contractions}
Contractions of three-dimensional Lie algebras over the real and complex fields
were exhaustively investigated in~\cite{NP2006}; however, to construct limit
transitions between realizations, it is necessary to align the bases of the
algebras with chains of nested subalgebras and construct contraction matrices
compatible with the chosen subalgebras. Therefore, we revisit this problem in
the present section.

Using existence criteria for contractions, we determine which complex
three-dimensional Lie algebras can be obtained by contracting the simple algebra
$\mathfrak{sl}(2, \mathbb{C})$.
The most effective criterion for this algebra is the \emph{unimodularity criterion}:
if there exists a contraction $\mathfrak{g}\to\mathfrak{g}_0$ and the algebra
$\mathfrak{g}$ is unimodular, then the algebra $\mathfrak{g}_0$ is also unimodular.
A Lie algebra $\mathfrak{g}$ is called unimodular if
$\operatorname{tr}(\operatorname{ad}_v)=0$ for all $v\in\mathfrak{g}$,
where $\operatorname{ad}_v(w)=[v,w]$.

The algebra $\mathfrak{sl}(2,\mathbb{C})$ is unimodular.
Checking the unimodularity condition for the remaining algebras in the
classification list~\eqref{classification3d} shows that among them, only the
algebras $\mathfrak{g}_{3.4}^{-1}$, $\mathfrak{g}_{3.1}$, and $3\mathfrak{g}_1$
are unimodular.

Indeed, for $\mathfrak{g}_0=\mathfrak{g}_{2.1}\oplus\mathfrak{g}_1$ we have
$\mathrm{tr}(\mathrm{ad}_{e_2})=-1\neq 0$;
for $\mathfrak{g}_{3.2}$,
$\mathrm{tr}(\mathrm{ad}_{e_3})=-2\neq 0$;
for $\mathfrak{g}_{3.3}$,
$\mathrm{tr}(\mathrm{ad}_{e_3})=-2\neq 0$;
and for the family $\mathfrak{g}_{3.4}^a$ we obtain
$\mathrm{tr}(\mathrm{ad}_{e_3}) = -(1+a)$,
so only the algebra $\mathfrak{g}_{3.4}^{-1}$, corresponding to the parameter
value $a = -1$, is unimodular.

Thus, among algebras non-isomorphic to $\mathfrak{sl}(2,\mathbb{C})$, the
necessary unimodularity condition is satisfied only by
$\mathfrak{g}_{3.4}^{-1}$, $\mathfrak{g}_{3.1}$, and $3\mathfrak{g}_1$.
Let us compare the dimensions of their derivation algebras:

\[
\begin{array}{c|cccc}
\mathfrak{g}
& \mathfrak{sl}(2,\mathbb{C})
& \mathfrak{g}_{3.4}^{-1}
& \mathfrak{g}_{3.1}
& 3\mathfrak{g}_1 \\ \hline
\dim\operatorname{Der}(\mathfrak{g})
& 3 & 4 & 6 & 9
\end{array}
\]

Under a proper contraction, the dimension of the derivation algebra increases
strictly. Hence, for these four algebras, the possible direction of proper
contractions is consistent with the ordering
\begin{equation}\label{contraction_sequence}
\mathfrak{sl}(2, \mathbb{C}) \to \mathfrak{g}_{3.4}^{-1} \to \mathfrak{g}_{3.1} \to 3\mathfrak{g}_1.
\end{equation}
This criterion rules out proper contractions in the reverse direction, but by
itself does not prove the existence of transitions in the direction of the arrows.
The existence of nontrivial transitions between adjacent algebras, as well as the
direct contraction $\mathfrak{sl}(2,\mathbb{C})\to\mathfrak{g}_{3.1}$, is
established by the matrices presented below.

In addition to transitions between adjacent algebras, the ordering above admits
the pairs
$\mathfrak{sl}(2, \mathbb{C}) \to \mathfrak{g}_{3.1}$,
$\mathfrak{sl}(2, \mathbb{C}) \to 3\mathfrak{g}_1$, and
$\mathfrak{g}_{3.4}^{-1} \to 3\mathfrak{g}_1$.
Their existence also follows from the matrix $U_3(\varepsilon)$ given below and
from the diagonal contraction to the Abelian algebra.

In this paper, we omit the commutative algebra and consider only three of the six
possible contraction pairs, namely
$\mathfrak{sl}(2, \mathbb{C}) \to \mathfrak{g}_{3.4}^{-1}$,
$\mathfrak{sl}(2, \mathbb{C}) \to \mathfrak{g}_{3.1}$, and
$\mathfrak{g}_{3.4}^{-1} \to \mathfrak{g}_{3.1}$.
It is for these contractions that we address the construction of limit realizations
and, in one case, of invariants.
Contraction to the Abelian algebra is well known and can be realized for any
algebra using a diagonal matrix with the contraction parameter on the diagonal.
Indeed, for $U(\varepsilon)=\varepsilon I$, we have
$[v,w]_\varepsilon=\varepsilon[v,w]\to 0$.
In the present work, primary attention is given to nontrivial limit realizations
and the corresponding invariants, so the Abelian algebra is excluded from consideration.

Consequently, up to isomorphism of the resulting algebra, the two nontrivial proper
contractions of $\mathfrak{sl}(2,\mathbb{C})$ listed above exhaust all possible cases.
This statement applies to classes of limit algebras; it is not a classification
of all contraction matrices or all limit transitions between their realizations.

For each of the pairs under consideration, we construct a corresponding matrix
as a product
\mbox{$U(\varepsilon)=I\cdot W(\varepsilon)$},
where $I$ is a constant nonsingular matrix and $W(\varepsilon)$ is a diagonal
matrix with exponents $0$ and $1$.
The matrix $I$ defines a preliminary change of basis in the initial algebra,
after which $W(\varepsilon)$ specifies an In\"on\"u--Wigner contraction.
Without such a change of basis, the matrix $W(\varepsilon)$ alone does not
necessarily define the same contraction.

To perform contractions exclusively via diagonal matrices (when a GIW contraction
exists), one must first align the bases of the initial and resulting algebras.
The problem of aligning bases for all representatives of non-isomorphic Lie
algebras of a given dimension is extraordinarily difficult.
A similar issue arises when constructing optimal subalgebra trees in order to
minimize computations needed to obtain all inequivalent realizations, as well as
when choosing optimal bases to construct the Hasse diagram of contractions.
For certain small sets of algebras, one can select aligned bases starting from
the algebra at the end of the chain, though this is not always possible.
Moreover, there exist pairs of algebras for which constructing a GIW contraction
requires simultaneous changes of basis in both the initial and target algebras.

The contraction $\mathfrak{sl}(2,\mathbb{C}) \to \mathfrak{g}_{3.4}^{-1}$ is
realized by the matrix $U_1(\varepsilon)=I_1\cdot W_1(\varepsilon)$, where
\begin{equation*}
  I_1=\begin{pmatrix} 1 & 0 & 0 \\ 0 & 0 & 1 \\ 0 & 1 & 0 \end{pmatrix}, \quad
  W_1(\varepsilon) =   \begin{pmatrix} \varepsilon & 0 & 0 \\ 0 & 1 & 0 \\ 0 & 0 & 1 \end{pmatrix},\quad
    U_1(\varepsilon) = \begin{pmatrix} \varepsilon & 0 & 0 \\ 0 & 0 & 1 \\ 0 & 1 & 0 \end{pmatrix}.
\end{equation*}

This matrix corresponds to the subalgebra $\langle e_2, e_3\rangle$.
Applying~\eqref{tens_str_eps}, we obtain the parameterized structure constants
that will be used in Section~\ref{Sec_param_realiz}:
\begin{gather}\label{c_sl2_g34}
\begin{array}{l}
  \tilde{c}_{12}^{3}(\varepsilon)=- \tilde{c}_{21}^{3}(\varepsilon)=2\varepsilon\to 0,\quad \varepsilon\to 0;\\
  \tilde{c}_{13}^{1}(\varepsilon)=-\tilde{c}_{31}^{1}(\varepsilon)=1\to 1,\quad \varepsilon\to 0;\\
  \tilde{c}_{23}^{2}(\varepsilon)=-\tilde{c}_{32}^{2}(\varepsilon)=-1\to -1,\quad \varepsilon\to 0.
  \end{array}
\end{gather}
The limit structure constants define the algebra $\mathfrak{g}_{3.4}^{-1}$.

Similarly, the contraction $\mathfrak{g}_{3.4}^{-1}\to\mathfrak{g}_{3.1}$ is
realized by the matrix $U_2(\varepsilon)=I_2\cdot W_2(\varepsilon)$, which
corresponds to the subalgebra $\langle e_1+e_2\rangle$:
\begin{equation*}
I_2=\begin{pmatrix} 2 & 1 & 0 \\ 0 & 1 & 0 \\ 0 & 0 & 1 \end{pmatrix},\quad
W_2(\varepsilon)=\begin{pmatrix} \varepsilon & 0 & 0 \\ 0 & 1 & 0 \\ 0 & 0 & \varepsilon \end{pmatrix},\quad
U_2(\varepsilon) =\begin{pmatrix} 2\varepsilon & 1 & 0 \\ 0 & 1 & 0 \\ 0 & 0 & \varepsilon \end{pmatrix},
\end{equation*}
\begin{gather}\label{c_g34_g31}
\begin{array}{l}
  \tilde{c}_{13}^{1}(\varepsilon)=-\tilde{c}_{31}^{1}(\varepsilon)=\varepsilon\to 0,\quad \varepsilon\to 0;\\
  \tilde{c}_{23}^{1}(\varepsilon)=-\tilde{c}_{32}^{1}(\varepsilon)=1\to 1,\quad \varepsilon\to 0;\\
  \tilde{c}_{23}^{2}(\varepsilon)=-\tilde{c}_{32}^{2}(\varepsilon)=-\varepsilon\to 0,\quad \varepsilon\to 0.
  \end{array}
\end{gather}
There exists an alternative matrix $\tilde{U}_2(\varepsilon)$ realizing the same
contraction $\mathfrak{g}_{3.4}^{-1}\to\mathfrak{g}_{3.1}$ and preserving the
same subalgebra $\langle e_1+e_2\rangle$, which has the form
\begin{equation*}
\tilde{I}_2=\begin{pmatrix} 1 & 1 & 0 \\ -1 & 1 & 0 \\ 0 & 0 & 1 \end{pmatrix},\quad
\tilde{W}_2(\varepsilon)=\begin{pmatrix} \varepsilon & 0 & 0 \\ 0 & 1 & 0 \\ 0 & 0 & \varepsilon \end{pmatrix},\quad
\tilde{U}_2(\varepsilon)=\begin{pmatrix} \varepsilon & 1 & 0 \\ -\varepsilon & 1 & 0 \\ 0 & 0 & \varepsilon \end{pmatrix},
\end{equation*}
\begin{gather}\label{c_g34_g31_alt}
\begin{array}{l}
  \tilde{c}_{13}^{2}(\varepsilon)=-\tilde{c}_{31}^{2}(\varepsilon)=\varepsilon^2\to 0,\quad \varepsilon\to 0;\\
  \tilde{c}_{23}^{1}(\varepsilon)=-\tilde{c}_{32}^{1}(\varepsilon)=1\to 1,\quad \varepsilon\to 0.\\
   \end{array}
\end{gather}

The contraction $\mathfrak{sl}(2,\mathbb{C}) \to \mathfrak{g}_{3.1}$ can be
realized by the matrix $U_3(\varepsilon)=I_3\cdot W_3(\varepsilon)$, which
preserves the subalgebra $\langle e_3\rangle$:
\begin{equation*}
  I_3=\begin{pmatrix} 0 & 1 & 0 \\ 1 & 0 & 0 \\ 0 & 0 & \frac{1}{2} \end{pmatrix},\quad
  W_3(\varepsilon)=\begin{pmatrix} \varepsilon & 0 & 0 \\ 0 & \varepsilon & 0 \\ 0 & 0 & 1 \end{pmatrix},\quad
  U_3(\varepsilon)=\begin{pmatrix} 0 & \varepsilon & 0 \\ \varepsilon & 0 & 0 \\ 0 & 0 & \frac{1}{2} \end{pmatrix},
\end{equation*}
\begin{gather}\label{c_sl2_g31}
\begin{array}{l}
  \tilde{c}_{12}^{2}(\varepsilon)=- \tilde{c}_{21}^{2}(\varepsilon)=-\varepsilon\to 0,\quad \varepsilon\to 0;\\
  \tilde{c}_{13}^{3}(\varepsilon)=-\tilde{c}_{31}^{3}(\varepsilon)=\varepsilon\to 0,\quad \varepsilon\to 0;\\
  \tilde{c}_{23}^{1}(\varepsilon)=-\tilde{c}_{32}^{1}(\varepsilon)=1\to 1,\quad \varepsilon\to 0.
  \end{array}
\end{gather}

\section{Parameterized realizations and their contractions}\label{Sec_param_realiz}

Section~1 described the algebraic method for constructing realizations, whereby
each transitive local realization of an algebra $\mathfrak{g}$ by vector fields
on an $m$-dimensional manifold $M$ is determined by a pair
$(\mathfrak{g},\mathfrak{h})$, where $\mathfrak{h}\subset\mathfrak{g}$ is a
subalgebra of codimension $m$ playing the role of the stationary subalgebra.

The practical algorithm for constructing a realization consists of the following steps:
\begin{enumerate}
  \item Choose a basis $\{e_1,\dots,e_n\}$ of the algebra $\mathfrak{g}$ such
        that the vectors $e_{m+1},\dots,e_n$ form a basis of the subalgebra
        $\mathfrak{h}$.

  \item Construct the matrices of the adjoint representation
        $(\ad_i)^k_{j} = c^k_{ij}$.

  \item In canonical coordinates of the second kind, compute the component matrix
        of the Maurer--Cartan form. To be specific, set
  \[
    g(x)=\exp(-x_1e_1)\cdots\exp(-x_ne_n),\qquad
    dg\,g^{-1}=\sum_{i,j=1}^{n}\omega^i{}_j(x)e_i\,dx_j.
  \]
  Then the $j$-th column of the matrix $\omega$ is given by
  \[
    \omega_{\cdot j}(x)=
    -\exp(-x_1\ad_{e_1})\cdots
     \exp(-x_{j-1}\ad_{e_{j-1}})\,e_j,
  \]
  where $e_j$ on the right-hand side is viewed as the coordinate column vector
  of the basis element. For $j=1$, the product of exponentials equals the identity
  matrix.

  \item Find the vector fields
  \[
    \Xi_j=-\sum_{i=1}^{n}(\omega^{-1})^i{}_j
    \frac{\partial}{\partial x_i},\qquad j=1,\dots,n.
  \]
  The chosen sign ensures the commutation relations $[\Xi_i,\Xi_j]=c^k_{ij}\Xi_k$.

  \item Project the vector fields onto the quotient space $G/H$ for the right
  action of the subgroup $H$ on $G$. In the chosen local coordinates,
  $x_{m+1},\dots,x_n$ are coordinates along the fibers. Once projectability of
  the fields is verified, their components along
  $\partial/\partial x_1,\dots,\partial/\partial x_m$ are evaluated on the
  section $x_{m+1}=\cdots=x_n=0$. It is projectability, rather than merely
  setting these coordinates to zero, that guarantees the validity of this procedure.
\end{enumerate}

In formula~\eqref{1-form}, for the coordinates $g=\exp(x_1e_1)\cdots\exp(x_ne_n)$,
the left-invariant form $g^{-1}dg$ is used. Its dual system consists of left-invariant
vector fields, which are not in general projectable onto $G/H$ under the right
action of $H$. Therefore, the steps above employ the compatible right-invariant form.
The matrix construction of invariant forms and vector fields is presented in~\cite{MMS2015}.

As an example, consider the Weyl--Heisenberg algebra with the sole nonzero
commutator $[e_2,e_3]=e_1$ and the subalgebra $\mathfrak h=\langle e_3\rangle$.
For $g=\exp(-x_1e_1)\exp(-x_2e_2)\exp(-x_3e_3)$, we have
\[
  \omega=
  \begin{pmatrix}
  -1&0&x_2\\
  0&-1&0\\
  0&0&-1
  \end{pmatrix},\qquad
  -\omega^{-1}=
  \begin{pmatrix}
  1&0&x_2\\
  0&1&0\\
  0&0&1
  \end{pmatrix}.
\]
Thus, on the group we obtain the fields
$\Xi_1=\partial_1$, $\Xi_2=\partial_2$,
$\Xi_3=x_2\partial_1+\partial_3$. Their first two components do not depend on
the fiber coordinate $x_3$, so after projecting onto $G/H$ we get
\[
  R(e_1)=\partial_1,\qquad
  R(e_2)=\partial_2,\qquad
  R(e_3)=x_2\partial_1.
\]
Direct verification yields $[R(e_2),R(e_3)]=R(e_1)$, with all remaining
commutators equal to zero.

The key idea in modifying the algorithm to find contractions of realizations is
to substitute the parameterized structure constants $\tilde{c}^{k}_{ij}(\varepsilon)$
from~\eqref{tens_str_eps} into the algorithm instead of the fixed structure
constants $c^k_{ij}$.
This leads to $\varepsilon$-dependent matrices of the adjoint representation
\begin{equation}\label{eq:adj_eps}
  (\ad_i)^k_{j}(\varepsilon)
  = \tilde{c}^{k}_{ij}(\varepsilon)
  = \bigl((U(\varepsilon))^{-1}\bigr)^k_{k'}\,
    \bigl(U(\varepsilon)\bigr)^{i'}_{i}\,
    \bigl(U(\varepsilon)\bigr)^{j'}_{j}\,
    c^{k'}_{i'j'},
\end{equation}
and hence to an $\varepsilon$-dependent Maurer--Cartan form matrix
$\omega_\varepsilon(x)$ and vector fields
\[
  \Xi_j(\varepsilon)=
  -\sum_{i=1}^{n}(\omega_\varepsilon^{-1})^i{}_j
  \frac{\partial}{\partial x_i}.
\]

Upon projection, we obtain a parameterized realization $R^\varepsilon$ of the
algebra $\mathfrak g_\varepsilon=(V,[\cdot,\cdot]_\varepsilon)$:
\[
  [R^\varepsilon(e_i),R^\varepsilon(e_j)]
  =\sum_k\tilde c^k_{ij}(\varepsilon)R^\varepsilon(e_k).
\]
For $\varepsilon\ne 0$, the map $U(\varepsilon)\colon\mathfrak g_\varepsilon\to\mathfrak g$
is an isomorphism. Therefore, a realization of the original fixed algebra
$\mathfrak g$ is obtained as $R^\varepsilon\circ U(\varepsilon)^{-1}$.
It is after this identification that the definition of weak equivalence of
realizations of a fixed algebra is applied.

By a limit transition between realizations, we mean the convergence of the
coefficients of the fields $R^\varepsilon(e_i)$ to the coefficients of the fields
$R^0(e_i)$ in common local coordinates under the conditions specified below.
A contraction of an abstract algebra alone does not uniquely determine a limit
transition between realizations: one must also specify a family of realizations,
coordinates, stationary subalgebras, and the mode of convergence.

Formula~\eqref{eq:adj_eps} admits a simple matrix interpretation:
\begin{equation*}
\ad_j(\varepsilon) = (U(\varepsilon))^{-1}
\bigl((U(\varepsilon))^{j'}_{j}\ad_{j'}\bigr)U(\varepsilon),
\end{equation*}
where the expression in parentheses is a linear combination of the initial
adjoint representation matrices with $\varepsilon$-dependent coefficients.

For commutators to pass to the limit, local $C^1$-convergence of the vector field
coefficients suffices. Pointwise convergence of the matrices $\omega_\varepsilon$
alone is not sufficient for this purpose. For prolongations to pass to the limit,
convergence of the corresponding number of derivatives is required.

In the matrix construction above, the convergence
$\tilde c^k_{ij}(\varepsilon)\to\hat c^k_{ij}$ implies local uniform convergence
of the matrix exponentials and their coordinate derivatives.
Since $\omega_0(0)=-I$, one can fix a common neighborhood of the origin in which
$\omega_\varepsilon$ are nonsingular for all sufficiently small $\varepsilon$.
In this neighborhood, the inverse matrices and their coordinate derivatives also
converge locally uniformly. With a compatible choice of projectable fields and
local coordinates on the quotient space, we obtain in the limit a realization
of the algebra $\mathfrak g_0$.

For a limit transition between invariant systems, one must additionally verify
the convergence of the systems themselves and of the prolonged generators; the
corresponding conditions are formulated in~\cite{Ivanova2007}.

Let $\mathfrak h\subset\mathfrak g$ be a stationary subalgebra defining a
transitive realization $R$ of the algebra $\mathfrak g$.
It corresponds to the family of subalgebras
$\mathfrak h_\varepsilon=U(\varepsilon)^{-1}\mathfrak h$ of the algebras
$\mathfrak g_\varepsilon$. If $\mathfrak h_\varepsilon\to\mathfrak h_0$ in the
Grassmannian, then $\mathfrak h_0$ is a subalgebra of $\mathfrak g_0$ of the
same dimension. Indeed, for convergent bases $u_i(\varepsilon)\to u_i(0)$, we have
\[
  [u_i(\varepsilon),u_j(\varepsilon)]_\varepsilon
  \longrightarrow
  [u_i(0),u_j(0)]_0\in\mathfrak h_0.
\]
The full limit of a family of subspaces need not exist, but every Grassmannian
limit point of it is a subalgebra. Moreover, diagonality of $U(\varepsilon)$ by
itself does not imply that an arbitrary subspace remains invariant. For instance,
$\operatorname{diag}(\varepsilon,1)\langle e_1+e_2\rangle  =\langle\varepsilon e_1+e_2\rangle.$
Preservation of a subspace requires compatibility with the weight decomposition.
Furthermore, the statement that a fixed subspace is a subalgebra for all
$\mathfrak g_\varepsilon$ and the statement $U(\varepsilon)\mathfrak h=\mathfrak h$
are distinct.

In all cases considered in this paper, the subalgebras from which the realizations
are constructed are listed in the corresponding tables, and their closure under
the parameterized brackets is verified separately.

For a transitive local realization $R$ with stationary subalgebra $\mathfrak h$,
the standard relation $\ker R=\operatorname{core}_{\mathfrak g}(\mathfrak h)$
holds, where $\operatorname{core}_{\mathfrak g}(\mathfrak h)$ is the largest
ideal of $\mathfrak g$ contained in $\mathfrak h$. Consequently, $R$ is faithful
if and only if $\mathfrak h$ contains no nonzero ideals of $\mathfrak g$.

For a regular family of orbit realizations with stationary subalgebras
$\mathfrak h_\alpha$, the kernel of the entire realization equals
$\bigcap_\alpha\operatorname{core}_{\mathfrak g}(\mathfrak h_\alpha)$.
Thus, faithfulness of each orbit restriction is sufficient, though not necessary,
for faithfulness of the entire family. Applying this well-known criterion to the
limit realization does not constitute a separate criterion for preservation of
faithfulness under contraction.

Under contraction, the situation may change: the limit subalgebra
$\mathfrak{h}_0\subset\mathfrak{g}_0$ may contain a nontrivial ideal of
$\mathfrak{g}_0$, even if $\mathfrak{h}$ contained no ideals of $\mathfrak{g}$.
This is because upon contraction, the algebra $\mathfrak{g}_0$ may possess a
larger set of ideals than $\mathfrak{g}$ (in particular, contraction can
increase the dimension of the center and decrease the dimension of the derived
subalgebra). The dimension of the derived subalgebra does not increase, since it
is the rank of the map defined by the Lie bracket. For the chain extended to the
Abelian algebra, these dimensions are $3\to 2\to 1\to 0$.
Nevertheless, in all cases considered below, we succeeded in constructing
contractions in which the limit stationary subalgebras contain no ideals, so the
corresponding limit realizations are faithful.

\subsection{Contractions of realizations of the algebra $\mathfrak{sl}(2,{\mathbb C})$}
Before constructing $\varepsilon$-realizations and examining their limit relations,
we list the realizations of the three algebras under study obtained by the direct
method.

List of inequivalent realizations of the algebra $\mathfrak{sl}(2,{\mathbb C})$:
\begin{equation}\label{real_sl}
\begin{array}{l}
  R_1\colon\partial_1,\ x_1\partial_1+x_2\partial_2,\ x_1^2\partial_1+2x_1x_2\partial_2+x_2\partial_3;\\
  R_2\colon\partial_1,\ x_1\partial_1 +x_2\partial_2,\ (x_1^2+x_2^2)\partial_1+2x_1x_2\partial_2; \\
  R_3\colon\partial_1,\ x_1\partial_1 +x_2\partial_2,\ x_1^2\partial_1+2x_1x_2\partial_2;\\
  R_4\colon\partial_1,\ x_1\partial_1,\ x_1^2\partial_1.
\end{array}
\end{equation}

List of inequivalent realizations of the algebra $\mathfrak{g}_{3.4}^{-1}$:
\begin{equation}\label{real_34}
\begin{array}{l}
  \mathfrak{R}_1\colon\partial_1,\ \partial_2,\ x_1\partial_1-x_2\partial_2+\partial_3;\\
  \mathfrak{R}_2\colon\partial_1,\ \partial_2,\ x_1\partial_1-x_2\partial_2; \\
  \mathfrak{R}_3\colon\partial_1,\ x_2\partial_1,\ x_1\partial_1+2x_2\partial_2.
\end{array}
\end{equation}

List of inequivalent realizations of the algebra $\mathfrak{g}_{3.1}$:
\begin{equation}\label{real_31}
\begin{array}{l}
  \mathcal{R}_1\colon\partial_1,\ \partial_2,\ x_2\partial_1+\partial_3;\\
  \mathcal{R}_2\colon\partial_1,\ \partial_2,\ x_2\partial_1+x_3\partial_2; \\
  \mathcal{R}_3\colon\partial_1,\ \partial_2,\ x_2\partial_1.
\end{array}
\end{equation}

If we directly apply a contraction matrix as a linear transformation of a Lie
algebra of vector fields, we may obtain an unfaithful realization.
The following example shows that this can occur for a particular choice of the
initial realization.
Indeed, consider the realization $\mathfrak{R}_1$ of the algebra
$\mathfrak{g}_{3.4}^{-1}$ and perform the linear transformation corresponding
to the contraction matrix $U_2(\varepsilon)$. Taking the limit as $\varepsilon\to 0$,
we obtain an unfaithful realization of the algebra $\mathfrak{g}_{3.1}$:
\[
\begin{array}{l}
R^{\varepsilon}_1(e_1)=2\varepsilon\partial_1\to 0,\\
R^{\varepsilon}_1(e_2)=\partial_1+\partial_2\to \partial_1+\partial_2,\\
R^{\varepsilon}_1(e_3)=\varepsilon(x_1\partial_1-x_2\partial_2+\partial_3)\to 0.
\end{array}
\]

Therefore, in order to use the contraction matrices already obtained to find
limit transitions between realizations, it is necessary to construct a
parameterized family of realizations for the parameterized family of Lie algebras
$\mathfrak{g}_{\varepsilon}$ with structure constants $\tilde{c}(\varepsilon)$.
To construct $\varepsilon$-realizations, we apply the modified algebraic method
described above.

Let us construct a parameterized transitive realization of $\mathfrak g_\varepsilon$
with respect to the zero subalgebra, using the parameterized structure constants
tensor from~\eqref{c_g34_g31}. After identifying
$\mathfrak g_\varepsilon\simeq\mathfrak g_{3.4}^{-1}$ for $\varepsilon\ne 0$, we
obtain a realization weakly equivalent to $\mathfrak{R}_1[\mathfrak{g}_{3.4}^{-1}]$.
This family converges to the realization $\mathcal{R}_1[\mathfrak{g}_{3.1}]$ as
$\varepsilon\to 0$:
\[
\begin{array}{l}
R^{\varepsilon}(e_1)=\partial_1\to \partial_1,\quad \varepsilon\to 0;\\
R^{\varepsilon}(e_2)=\partial_2\to \partial_2  ,\quad \varepsilon\to 0;\\
R^{\varepsilon}(e_3)=(\varepsilon x_1+x_2)\partial_1-\varepsilon x_2\partial_2+\partial_3\to x_2\partial_1+\partial_3,\quad \varepsilon\to 0.
\end{array}
\]

Consider one more example of constructing a parameterized realization for the
algebra $\mathfrak{sl}(2, \mathbb C)$ contracting to $\mathfrak{g}_{3.1}$.
For the zero subalgebra $\mathfrak{h}_0=\{0\}$ and the contraction matrix
$U_3(\varepsilon)$, we obtain a family of realizations converging to
$\mathcal{R}_1[\mathfrak{g}_{3.1}]$:
\[
\begin{array}{l}
R^{\varepsilon}_1(e_1)=\partial_1\to \partial_1,\quad \varepsilon\to 0;\\
R^{\varepsilon}_1(e_2)={e}^{-\varepsilon x_1}\partial_2\to \partial_2,\quad \varepsilon\to 0;\\
R^{\varepsilon}_1(e_3)=x_2{e}^{\varepsilon x_1}\partial_1-\tfrac12 {e}^{\varepsilon x_1}\varepsilon x_2^2\partial_2+{e}^{\varepsilon x_1}\partial_3\to x_2\partial_1+\partial_3,\quad \varepsilon\to 0.
\end{array}
\]

Similarly, applying the corresponding contraction matrices and the modified
algebraic method to the other pairs of algebras yields the results compiled in
the tables and analyzed below.

Each table below specifies the subalgebra $\mathfrak{h}$ and complementary set
of elements $\mathfrak{\bar{h}}$ used to construct each parameterized realization,
as well as the corresponding canonical realizations from the
lists~\eqref{real_sl},~\eqref{real_34}, and~\eqref{real_31} to which they are
equivalent.
In non-obvious cases, the corresponding biholomorphisms and changes of basis are
written out explicitly after the tables.
Equivalence for $\varepsilon\ne 0$ is understood after identifying
$\mathfrak g_\varepsilon$ with the corresponding initial algebra.

\begin{table}[H]\renewcommand{\arraystretch}{1.1}
\centering
\caption{Realizations corresponding to the contraction $U_1(\varepsilon)\colon\mathfrak{sl}(2,\mathbb{C})\to\mathfrak{g}_{3.4}^{-1}$.}\label{tab:sl2_to_g34_reals}
\begin{tabular}{|c|c|l|l|}
    \hline
     && \multicolumn{1}{c|}{$\mathfrak{sl}(2, \mathbb{C})$} & \multicolumn{1}{c|}{$\mathfrak{g}_{3.4}^{-1}$} \\
    $\mathfrak{h}$ &$\mathfrak{\bar{h}}$& \multicolumn{1}{c|}{$\varepsilon$-realization}  & \multicolumn{1}{c|}{Limit realization}  \\
    \hline
    $\langle0\rangle$ &$\langle e_1,\ e_2,\ e_3\rangle$&
    $R_1\sim\ \begin{aligned}
    &\partial_1,\\
    &\varepsilon x_1^2\partial_1 +(1 - 2\varepsilon x_1x_2)\partial_2 + 2\varepsilon x_1 \partial_3,\\
    &x_1\partial_1 - x_2\partial_2 + \partial_3;
    \end{aligned}$
    &$\mathfrak{R}_1\sim\ \begin{aligned}
    &\partial_1,\\
    &\partial_2,\\
    &x_1 \partial_1 - x_2 \partial_2 + \partial_3;
    \end{aligned}$ \\
    \hline
    $\langle e_3 \rangle$ &$\langle e_1, e_2 \rangle$&
    $R_2\sim\ \begin{aligned}
      &\partial_1, \\
      &\varepsilon x_1^2\partial_1 +(1 - 2\varepsilon x_1x_2)\partial_2, \\
      &x_1\partial_1 - x_2\partial_2;
    \end{aligned}$
    &
    $\mathfrak{R}_2\sim\ \begin{aligned}
      & \partial_1, \\
      & \partial_2, \\
      & x_1 \partial_1 - x_2 \partial_2;
    \end{aligned}$ \\
    \hline
    $\langle e_2 - e_1 \rangle$  &$\langle e_2, -e_3 \rangle$&
$R_2\sim\ \begin{aligned}
&(\varepsilon x_1^2 + {e}^{2x_2})\partial_1+ 2\varepsilon x_1 \partial_2, \\
&\partial_1, \\
&-\!x_1\partial_1\!-\!\partial_2;
\end{aligned}$
    &
$\mathfrak{R}_3\sim\ \begin{aligned}
&{e}^{2x_2}\partial_1, \\
&\partial_1, \\
&-\!x_1\partial_1\!-\!\partial_2.
\end{aligned}$\\
    \hline
    \end{tabular}
\end{table}

The equivalence of the parameterized realization $R_1^\varepsilon$ for
$\varepsilon\ne 0$ to the canonical case $R_1$ from~\eqref{real_sl} is established
by the biholomorphism $\tilde{x}_1=x_1$, $\tilde{x}_2={e}^{x_3}$,
$\tilde{x}_3=\varepsilon x_2{e}^{x_3}$ and the change of basis
$\tilde{e}_1 = e_1$, $\tilde{e}_2 = e_3$, $\tilde{e}_3 =\frac{1}{\varepsilon} e_2$.

The equivalence of the realization $R_2^\varepsilon$ constructed from the
subalgebra $\langle e_3 \rangle$ for $\varepsilon\ne 0$ to the canonical case
$R_2$ from~\eqref{real_sl} is established by the biholomorphism
$\tilde{x}_1=\varepsilon x_1-\frac{1}{2x_2}$, $\tilde{x}_2=\frac{i}{2x_2}$
and the change of basis $\tilde{e}_1 = e_1$, $\tilde{e}_2 = e_3$, $\tilde{e}_3 = e_2$.

The equivalence of the third parameterized realization $R_2^\varepsilon$ for
$\varepsilon\ne 0$ to the canonical case $R_2$ from~\eqref{real_sl} is established
by the biholomorphism $\tilde{x}_1=x_1$,
$\tilde{x}_2=\frac{i}{\sqrt{\varepsilon}}{e}^{x_2}$ and the change of basis
$\tilde{e}_1 = e_1$, $\tilde{e}_2 = \frac{1}{\varepsilon} e_3$,
$\tilde{e}_3 = e_2$.

To reduce the limit realization obtained from the subalgebra $\langle e_2 - e_1 \rangle$
to the canonical form $\mathfrak{R}_3$ of the algebra $\mathfrak{g}_{3.4}^{-1}$
from~\eqref{real_34}, the biholomorphism $\tilde{x}_1 = x_1$,
$\tilde{x}_2 = e^{2x_2}$ suffices.

\begin{remark}\label{remark_ekviv_to_nonekviv}
  The last two $\varepsilon$-realizations of the algebra
  $\mathfrak{sl}(2, \mathbb{C})$ in Table~\ref{tab:sl2_to_g34_reals} are
  equivalent for every $\varepsilon\ne 0$, yet in the limit they yield two
  distinct faithful and inequivalent realizations of the algebra
  $\mathfrak{g}_{3.4}^{-1}$.
  They correspond to the same contraction matrix, but are constructed from
  different one-dimensional subalgebras.
\end{remark}

\begin{table}[H]\renewcommand{\arraystretch}{1.1}
\centering
\caption{Realizations corresponding to the contraction $\tilde{U}_2(\varepsilon)\colon\mathfrak{g}_{3.4}^{-1}\to\mathfrak{g}_{3.1}$.}\label{tab:g34_to_g31_reals}
\begin{tabular}{|c|c|l|l|}
    \hline
     && \multicolumn{1}{c|}{$\mathfrak{g}_{3.4}^{-1}$} & \multicolumn{1}{c|}{$\mathfrak{g}_{3.1}$} \\
    $\mathfrak{h}$ &$\bar{\mathfrak{h}}$& \multicolumn{1}{c|}{$\varepsilon$-realization}  & \multicolumn{1}{c|}{Limit realization}  \\
    \hline
    $\langle 0 \rangle$ &$\langle e_1, e_2, e_3 \rangle$&
    $\mathfrak{R}_1\sim\ \begin{aligned}
    &\partial_1, \\
    &\partial_2, \\
    &x_2\partial_1 + \varepsilon^2 x_1\partial_2 + \partial_3;
    \end{aligned}$
    &
    $\mathcal{R}_1\sim\ \begin{aligned}
    &\partial_1, \\
    &\partial_2, \\
    &x_2\partial_1 + \partial_3;
    \end{aligned}$ \\
    \hline
    $\begin{aligned}
    &\langle e_3 - \alpha e_2 \rangle,\\
    &\alpha\neq 0,\\
    &x_3:=\alpha
    \end{aligned}$
    &$\langle e_1, e_2 \rangle$&
    $\mathfrak{R}_2\sim\ \begin{aligned}
    &\partial_1, \\
    &\partial_2, \\
    &x_2\partial_1 + (\varepsilon^2 x_1 + \alpha)\partial_2;
    \end{aligned}$
    &
    $\mathcal{R}_2\sim\
    \begin{aligned}
    &\partial_1, \\
    &\partial_2, \\
    &x_2\partial_1 + \alpha\partial_2;
    \end{aligned}$ \\
    \hline
    $\langle e_3 \rangle$ &$\langle e_1, e_2 \rangle$&
    $\mathfrak{R}_2\sim\ \begin{aligned}
    &\partial_1, \\
    &\partial_2, \\
    &x_2\partial_1 + \varepsilon^2 x_1\partial_2;
    \end{aligned}$
    &
    $\mathcal{R}_3\sim\ \begin{aligned}
    &\partial_1, \\
    &\partial_2, \\
    &x_2\partial_1.
    \end{aligned}$ \\
    \hline
    \end{tabular}
\end{table}

The equivalence between the parameterized realization $\mathfrak{R}_1^\varepsilon$
($\varepsilon\neq 0$) and the canonical realization $\mathfrak{R}_1$ is
established by the biholomorphism
$\tilde{x}_1=\varepsilon x_1+x_2$, $\tilde{x}_2=x_2-\varepsilon x_1$,
$\tilde{x}_3=\varepsilon x_3$ and the change of basis
$\tilde{e}_1=\frac 12(e_2+\frac{1}{\varepsilon} e_1)$,
$\tilde{e}_2=\frac 12(e_2-\frac{1}{\varepsilon} e_1)$,
$\tilde{e}_3=\frac{1}{\varepsilon} e_3$.

The equivalence between both realizations $\mathfrak{R}_2^\varepsilon$ and the
canonical realization $\mathfrak{R}_2$ is established by the biholomorphism
$\tilde{x}_1=\varepsilon x_1+x_2+\frac{\alpha}{\varepsilon}$,
$\tilde{x}_2={x_2}-\varepsilon x_1-\frac{\alpha}{\varepsilon}$
and the same change of basis as in the previous case.

\begin{remark}
  To obtain the intransitive realization
  $\mathcal{R}_2 = \langle \partial_1, \partial_2, x_2\partial_1 + x_3\partial_2 \rangle$
  from the list~\eqref{real_31}, we interpret the continuous parameter $\alpha$
  as the third coordinate variable $x_3$, which is an invariant of this
  realization constructed on a two-dimensional manifold.

  As in Remark~\ref{remark_ekviv_to_nonekviv}, the last two $\varepsilon$-realizations
  are equivalent to each other, yet in the limit they yield inequivalent realizations.
\end{remark}

\begin{table}[H]\renewcommand{\arraystretch}{1.1}
\centering
\caption{Realizations corresponding to the contraction $U_3(\varepsilon)\colon\mathfrak{sl}(2,\mathbb{C})\to\mathfrak{g}_{3.1}$.}\label{tab:sl2_to_g31_reals}
\begin{tabular}{|c|c|l|l|}
    \hline
     && \multicolumn{1}{c|}{$\mathfrak{sl}(2, \mathbb{C})$} & \multicolumn{1}{c|}{$\mathfrak{g}_{3.1}$} \\
    $\mathfrak{h}$ &$\bar{\mathfrak{h}}$& \multicolumn{1}{c|}{$\varepsilon$-realization}  & \multicolumn{1}{c|}{Limit realization}  \\
    \hline
    $\langle 0 \rangle$ &$\langle e_1, e_2, e_3 \rangle$&
    $R_4\sim\ \begin{aligned}
    &\partial_1, \\
    &{e}^{-2\varepsilon x_1}\partial_2, \\
    &{e}^{2\varepsilon x_1}(x_2\partial_1 - \varepsilon x_2^2\partial_2 + \partial_3);
    \end{aligned}$
    &
    $\mathcal{R}_1\sim\ \begin{aligned}
    &\partial_1, \\
    &\partial_2, \\
    &x_2\partial_1 + \partial_3;
    \end{aligned}$ \\
    \hline
    $\begin{aligned}
    &\langle e_2 + \alpha e_3 \rangle,\\
    &\alpha\neq 0,\\
    &x_3:=\alpha
    \end{aligned}$
    &$\langle e_1, e_2 \rangle$&
    $R_2\sim\ \begin{aligned}
    &\partial_1, \\
    &{e}^{-2\varepsilon x_1}\partial_2, \\
    &{e}^{2\varepsilon x_1} (x_2\partial_1 + (\alpha - \varepsilon x_2^2)\partial_2);
    \end{aligned}$
    &
    $\mathcal{R}_2\sim\ \begin{aligned}
    &\partial_1, \\
    &\partial_2, \\
    &x_2\partial_1 + \alpha\partial_2;
    \end{aligned}$ \\
    \hline
    $\langle e_2 \rangle$ &$\langle e_1, e_3 \rangle$&
    $R_3\sim\ \begin{aligned}
    &\partial_1, \\
    &{e}^{2\varepsilon x_1}\partial_2, \\
    &-{e}^{-2\varepsilon x_1}(x_2\partial_1 + \varepsilon x_2^2\partial_2);
    \end{aligned}$
    &
    $\mathcal{R}_3\sim\ \begin{aligned}
    &\partial_1, \\
    &\partial_2, \\
    &x_2\partial_1;
    \end{aligned}$ \\
    \hline
    \end{tabular}
\end{table}

In all three rows of Table~\ref{tab:sl2_to_g31_reals}, establishing the
equivalences for the realizations in the third column uses the change of basis
$\tilde{e}_1=e_2$, $\tilde{e}_2=\frac{1}{2\varepsilon}e_1$,
$\tilde{e}_3=\frac{1}{\varepsilon}e_3$ together with nonsingular coordinate
transformations.
The equivalence of the parameterized realization in the first row to the canonical
realization $R_4$ is established by the biholomorphism
$\tilde{x}_1=x_2 {e}^{2\varepsilon x_1}$, $\tilde{x}_2={e}^{2\varepsilon x_1}$,
$\tilde{x}_3=\varepsilon x_3$ and the basis change $\tilde{e}_1=e_2$,
$\tilde{e}_2=\frac{1}{2\varepsilon} e_1$, $\tilde{e}_3=\frac{1}{\varepsilon} e_3$.

The parameterized realization $R_2^{\varepsilon}$ is equivalent to the canonical
realization $R_2$ after the biholomorphic change of variables
$\tilde{x}_1=x_2 {e}^{2\varepsilon x_1}$,
$\tilde{x}_2=-\left(\frac{\alpha}{\varepsilon}\right)^{-\frac 12}{e}^{-2\varepsilon x_1}$.

The equivalence $R_3^{\varepsilon}\sim R_3$ is established by the biholomorphism
$\tilde{x}_1=x_2 {e}^{2\varepsilon x_1}$, $\tilde{x}_2={e}^{2\varepsilon x_1}$.

\begin{remark}
  In the case of the contraction $\mathfrak{sl}(2,\mathbb{C})\to\mathfrak{g}_{3.1}$,
  no splitting of families of equivalent realizations into inequivalent ones occurs
  in the limit; namely, the parameterized realizations of
  $\mathfrak{sl}(2, \mathbb{C})$ in Table~\ref{tab:sl2_to_g31_reals} constructed
  from the subalgebras $\langle e_2+\alpha e_3\rangle$ and $\langle e_2\rangle$
  are inequivalent, and in the limit they likewise pass into inequivalent
  canonical realizations of the algebra $\mathfrak{g}_{3.1}$.
\end{remark}

\section{Differential invariants and invariant systems}\label{Sec_invariants}

The problem of describing differential equations invariant under a given Lie
algebra reduces to finding the differential invariants of its realizations.
On regular domains, invariant systems can be specified by relations among
differential invariants of the appropriate order.

The classical theory of differential invariants originates in the works of
Lie~\cite{Lie1884, Lie1893} and Tresse~\cite{Tresse1894, Tresse1896};
in this section, we rely on the systematic exposition of equivalence and
invariants in Olver's monograph~\cite{Olver1993} and on the method of moving
coframes proposed by Fels and Olver~\cite{FelsOlver1998,FelsOlver1999}.
For transitions between different realizations, multiple dependent functions,
and multiple independent variables, see~\cite{FY, Yehorchenko1996}.
A generalization of Lie's classical theorem on the basis of differential
invariants of one-parameter local transformation groups to an arbitrary number
of independent and dependent variables, as well as a method for constructing a
complete set of functionally independent invariants via a single quadrature and
differentiation using a known universal invariant, is given in~\cite{PopovychBoyko2002},
which also considers applications of first-order differential invariants to
Riccati-type systems.

Consider the case of one independent variable $t$ and two dependent variables
$x, y$. Suppose a realization of a Lie algebra ${\mathfrak g}$ is given by
holomorphic basis vector fields in a neighborhood of a three-dimensional complex
manifold $M$:
\begin{gather}\label{algebra_basis}
R(e_i) = \xi_i(t,x,y)\p_t + \eta_i(t,x,y)\p_x
+ \zeta_i(t,x,y)\p_y,\quad i=1,\dots, n.
\end{gather}
For brevity in this section, we denote the realizations of the basis elements
and their prolongations by $e_i$ and $e_i^{(k)}$ instead of $R(e_i)$ and
$R^{(k)}(e_i)$.
The local action of the algebra $\mathfrak g$ on the jet space $M^{(k)}$ is
defined by the $k$-th prolonged operators
\begin{gather}\label{algebra_basis_prolongated}
e_i^{(k)} = e_i + \sum_{l=1}^{k}
\left(\eta_i^{l}(t,x_{(l)},y_{(l)})\p_{x^{(l)}}
+ \zeta_i^{l}(t,x_{(l)},y_{(l)})\p_{y^{(l)}}\right),
\end{gather}
where $x_{(l)} = (x, \dot x, \dots, x^{(l)})$,
$y_{(l)} = (y, \dot y, \dots, y^{(l)})$,
and the coefficients $\eta_i^l, \zeta_i^l$ are computed via the standard
Lie--Ovsiannikov formulas~\cite{Ovsiannikov1982} using the total derivative
operator $D_t$.

\begin{definition}
A \emph{differential invariant} of order~$k$ of the algebra~$\mathfrak g$ is a
holomorphic function $I(t,x_{(k)},y_{(k)})$ satisfying the determining system
$e_i^{(k)} I = 0$, $i=1,\dots,n$.
\end{definition}

Analogously to the rank of a realization, one can consider the ranks of
prolonged realizations. Let $r_k$ denote the maximal rank of the matrix formed
by the coefficients of the prolonged operators $e_i^{(k)}$; henceforth, we
consider domains on which the corresponding rank is constant.

For the faithful local holomorphic realizations considered here, the nondecreasing
sequence of maximal ranks $\{r_k\}$ stabilizes at the value $n$. Hence, there
exists $\nu=\min\{k\in\mathbb N_0\mid r_k=n\}$ such that
$r_\nu=r_{\nu+1}=\dots=n$.

The number of functionally independent invariants of order~$k$ on a regular
domain is given by the formula
$d_k = 2k + 3 - r_k.$

A maximal set ${\rm I}_k$ of functionally independent invariants of order at
most~$k$ forms a \emph{universal differential invariant} of order~$k$.

On an appropriate regular domain, a finite set of differential invariants
together with an invariant differentiation operator locally generates higher-order
invariants by means of functional operations and differentiation. For a single
independent variable, such an operator has the form $\delta=\lambda D_t$.

The coefficient function $\lambda(t,x_{(\nu)},y_{(\nu)})$ is determined from the
commutation condition $[e_i^{(\infty)}, \delta] = 0$:
\begin{gather*}
e_i^{(\nu)}(\lambda) - \lambda D_t(\xi_i) = 0,\quad i=1,\dots,n.
\end{gather*}

Once a complete set of functionally independent second-order differential
invariants $I_1,\dots,I_d$ is found, a regular invariant system of two
second-order ordinary differential equations can be locally specified by the
relations
\[
\Phi_1(I_1,\dots,I_d)=0,\qquad
\Phi_2(I_1,\dots,I_d)=0.
\]
The solvability of such a system for the highest-order derivatives (normal form)
is determined by the ranks of the prolongations.

\begin{proposition}
A Lie algebra of vector fields~$\mathfrak g$ admits a local normal regular
invariant system
\begin{gather*}
\ddot x = F(t,x,y,\dot x,\dot y),\quad
\ddot y = G(t,x,y,\dot x,\dot y)
\end{gather*}
if and only if $r_1 = r_2$ on the corresponding regular domain.
\end{proposition}

In addition to the regular invariant systems mentioned above, there may also
exist singular invariant systems described using Lie determinants, though we do
not consider such cases in the present work.

\subsection{Parameterized invariants of a generating realization of the algebra $\mathfrak{sl}(2, \mathbb{C})$}

Consider an example of constructing an invariant system of the form~\eqref{gen_sys}
for the first $\varepsilon$-realization of the algebra $\mathfrak{sl}(2, \mathbb{C})$
in Table~\ref{tab:sl2_to_g34_reals}, corresponding to the zero subalgebra:
$$
\partial_1,\ \varepsilon x_1^2\partial_1
+(1 - 2\varepsilon x_1x_2)\partial_2
+ 2\varepsilon x_1 \partial_3,\
x_1\partial_1 - x_2\partial_2 + \partial_3.
$$

To simplify the computations, we apply the change of coordinates
\begin{equation}\label{eq_var_change}
\tilde{x}_1=x_1,\quad
\tilde{x}_2={e}^{x_3},\quad
\tilde{x}_3=\varepsilon x_2 {e}^{x_3}.
\end{equation}

Its Jacobian equals
\[
\det\frac{\partial(\tilde{x}_1,\tilde{x}_2,\tilde{x}_3)}
{\partial(x_1,x_2,x_3)}
=-\varepsilon e^{2x_3}.
\]
Hence, for $\varepsilon\ne 0$, this is a local biholomorphism. It is not globally
injective on $\mathbb C^3$, since the exponential function is not globally
injective. To bring the vector fields to canonical form, the change of coordinates
is accompanied by the change of basis
\[
\widetilde e_1=e_1,\qquad
\widetilde e_2=e_3,\qquad
\widetilde e_3=\frac{1}{\varepsilon}e_2.
\]

As a result, we obtain the canonical case $R_1$ from~\eqref{real_sl}:
\begin{equation*}
\partial_{\tilde{x}_1}, \quad
\tilde{x}_1\partial_{\tilde{x}_1}
+\tilde{x}_2\partial_{\tilde{x}_2}, \quad
\tilde{x}_1^2\partial_{\tilde{x}_1}
+2\tilde{x}_1\tilde{x}_2\partial_{\tilde{x}_2}
+\tilde{x}_2\partial_{\tilde{x}_3}.
\end{equation*}

Fix the independent variable $t$ and the two dependent variables as
$t:=\tilde{x}_1$, $x:=\tilde{x}_2$, $y:=\tilde{x}_3$.
Throughout this example, we consider the regular domain $x\ne 0$. Denote the
canonical generators by
\[
X_1=\partial_t,\qquad
X_2=t\partial_t+x\partial_x,\qquad
X_3=t^2\partial_t+2tx\partial_x+x\partial_y.
\]

For this realization, the rank stabilizes already at the zeroth prolongation:
on the domain $x\ne 0$, we have $r_0=r_1=r_2=3$.

Constructing the first prolongations of the operators and solving the determining
system yields two absolute first-order differential invariants:
\[
I_1= y-\frac{1}{2}\dot{x},\qquad
J_1 =\dot{x}y-\dot{y}x-y^2.
\]
One directly verifies that
\[
\operatorname{pr}^{(1)}X_i(I_1)=0,\qquad
\operatorname{pr}^{(1)}X_i(J_1)=0,\qquad i=1,2,3.
\]
Moreover,
\[
\det\frac{\partial(I_1,J_1)}
{\partial(\dot x,\dot y)}
=\frac{x}{2}\ne 0.
\]
Since $d_1=5-r_1=2$, the invariants $I_1,J_1$ form a complete local functionally
independent set on the first-order jet space.

For this realization, an invariant differentiation operator is $\delta=xD_t$.
Indeed,
\[
X_1(x)=0,\qquad X_2(x)=x,\qquad X_3(x)=2tx.
\]
Thus $x$ is not an absolute invariant: on the domain $x\ne 0$, it is a relative
invariant with multipliers $(0,1,2t)$. These multipliers define a local cocycle
rather than a constant character of the group $\mathrm{SL}(2,\mathbb C)$. Since
for the coefficients $\xi_i$ of $\partial_t$ we have
\[
X_i(x)=xD_t(\xi_i),\qquad i=1,2,3,
\]
it follows that $[\operatorname{pr}^{(\infty)}X_i,xD_t]=0$. At $x=0$, the
operator $xD_t$ degenerates, so the condition $x\ne 0$ is essential.

Applying $\delta$ to $I_1$ and $J_1$ produces second-order absolute invariants:
\[
I_2=x\left(\dot{y}-\frac{1}{2}\ddot{x}\right),\qquad
J_2=x\left(\ddot{x}y-x\ddot{y}-2y\dot{y}\right),
\]
for which
\[
\det\frac{\partial(I_2,J_2)}
{\partial(\ddot x,\ddot y)}
=\frac{x^3}{2}\ne 0.
\]
Hence $I_1,J_1,I_2,J_2$ are functionally independent. Since $d_2=7-r_2=4$, they
form a complete local set of differential invariants of order at most two.

We also note that for this realization, higher-order invariants are locally
generated from $I_1,J_1$ by repeated application of $\delta$ and functional
operations: at each subsequent order, the two new derivatives
$\delta^{k-1}I_1$ and $\delta^{k-1}J_1$ are independent with respect to the
highest-order derivatives for $x\ne 0$.

Since the ranks of the first and second prolongations coincide for the realization
$R_1$, the corresponding regular invariant system of Newtonian type can be
written in normal form. Consider first the absolute invariant equations
\[
I_2=F(I_1,J_1),\qquad J_2=G(I_1,J_1),
\]
where $F$ and $G$ are arbitrary local holomorphic functions. Solving for the
highest-order derivatives yields the system
\begin{equation}\label{eq:sys_R1_explicit}
\begin{cases}
    \ddot{x}=2\dot{y}-\frac{2}{x}F(I_1, J_1), \\
    \ddot{y}=-\frac{2y}{x^2}F(I_1, J_1)
    -\frac{1}{x^2}G(I_1, J_1),
\end{cases}
\end{equation}
which is defined on the domain $x\ne 0$.

To clarify the nature of its invariance, denote the left-hand sides of
system~\eqref{eq:sys_R1_explicit} by
\[
H_1=\ddot x-2\dot y+\frac{2F}{x},\qquad
H_2=\ddot y+\frac{2yF}{x^2}+\frac{G}{x^2}.
\]
Then
\[
I_2-F=-\frac{x}{2}H_1,\qquad
J_2-G=xyH_1-x^2H_2.
\]
The determinant of the transition matrix between the pairs of residuals
$(H_1,H_2)$ and $(I_2-F,J_2-G)$ equals $x^3/2$. Thus, on the domain $x\ne 0$,
they define the same invariant set, although $H_1$ and $H_2$ themselves need not
be individual absolute invariants.

In the real case, the resulting class of systems reproduces the corresponding
known class from~\cite{Haponova2006}. For comparison, we use the invariants
given there:
\[
\omega_1=2y-\dot x,\qquad
\omega_2=4x\dot y-\dot x^2.
\]
The relation between the two sets of invariants is
\[
I_1=\frac{\omega_1}{2},\qquad
J_1=-\frac{\omega_1^2+\omega_2}{4}.
\]

In particular, in the normalization of~\cite{Haponova2006}, the corresponding
system can be written as
\[
\begin{cases}
\displaystyle
\ddot x=2\dot y+\frac{1}{x}F_{26}(\omega_1,\omega_2),\\[5pt]
\displaystyle
\ddot y=\frac{y}{x^2}F_{26}(\omega_1,\omega_2)
+\frac{1}{x^2}G_{26}(\omega_1,\omega_2).
\end{cases}
\]
It is equivalent to system~\eqref{eq:sys_R1_explicit} upon relabeling the
arbitrary functions via
\[
\begin{aligned}
F_{26}(\omega_1,\omega_2)
&=-2F\left(\frac{\omega_1}{2},
-\frac{\omega_1^2+\omega_2}{4}\right),\\
G_{26}(\omega_1,\omega_2)
&=-G\left(\frac{\omega_1}{2},
-\frac{\omega_1^2+\omega_2}{4}\right).
\end{aligned}
\]
Thus, the construction of the canonical invariant system~\eqref{eq:sys_R1_explicit}
reproduces the known result~\cite{Haponova2006}; its parameterization and limit
transition are considered below.

\subsection{Construction of parameterized invariant systems}

To construct an $\varepsilon$-parameterized invariant system, we apply the
transformation locally inverse to~\eqref{eq_var_change}, written in terms of the
variables $t=\tilde{x}_1$, $x=\tilde{x}_2$, $y=\tilde{x}_3$ and the variables
$T:=x_1$ (independent variable), $X:=x_2$ and $Y:=x_3$ (dependent variables):
\begin{equation}\label{eq_inverse_var_change}
T=t,\qquad
X=\frac{y}{\varepsilon\,x},\qquad
Y=\ln x.
\end{equation}

Here $\ln x$ denotes a chosen local holomorphic branch of the logarithm on the
domain $x\ne 0$. In the real setting, this substitution corresponds to the
domain $x>0$. The transformation is defined only for $\varepsilon\ne 0$; therefore,
the limit transition will be performed after rewriting the system in the common
coordinates $(T,X,Y)$. Local biholomorphic equivalence of realizations for
$\varepsilon\ne 0$ does not by itself establish equivalence at $\varepsilon=0$.

Using $x=e^{Y}$ and $y=\varepsilon e^{Y}X$, and denoting $X'=\frac{dX}{dT}$,
$Y'=\frac{dY}{dT}$, $X''=\frac{d^2X}{dT^2}$, $Y''=\frac{d^2Y}{dT^2}$,
since $T=t$, derivatives with respect to $t$ coincide with derivatives with
respect to $T$, and we obtain
\begin{equation*}
\dot x=e^{Y}Y', \quad
\ddot x=e^{Y}\bigl(Y''+Y'^2\bigr),\quad
\dot y=\varepsilon e^{Y}\bigl(XY'+X'\bigr), \quad
\ddot y=\varepsilon e^{Y}
\bigl(X\bigl(Y''+Y'^2\bigr)+2Y'X'+X''\bigr).
\end{equation*}

Let $I_\varepsilon$ and $J_\varepsilon$ denote the expressions for the invariants
$I_1$ and $J_1$ from the preceding subsection after passing to the coordinates
$(T,X,Y)$. They take the form
\begin{equation*}
{I_\varepsilon}=
e^{Y}\Bigl(\varepsilon X-\tfrac12 Y'\Bigr),\quad
{J_\varepsilon}
=-\varepsilon e^{2Y}\bigl(X'+\varepsilon X^2\bigr).
\end{equation*}

For each fixed $\varepsilon\ne 0$, the functions $I_\varepsilon$ and
$J_\varepsilon$ are absolute differential invariants of the parameterized
realization. However, $J_\varepsilon\to 0$ as $\varepsilon\to 0$, so the limit of
the pair $(I_\varepsilon,J_\varepsilon)$ does not provide two functionally
independent invariants. To obtain a nontrivial limit, consider the normalized
invariants
\[
A_\varepsilon=-2I_\varepsilon
=e^Y\bigl(Y'-2\varepsilon X\bigr),\qquad
B_\varepsilon=-\frac{J_\varepsilon}{\varepsilon}
=e^{2Y}\bigl(X'+\varepsilon X^2\bigr).
\]
For each fixed $\varepsilon\ne 0$, the factors $-2$ and $-1/\varepsilon$ are
constants with respect to the jet-space variables, so $A_\varepsilon$ and
$B_\varepsilon$ remain absolute invariants.

In the common jet-space coordinates, taking $\varepsilon\to 0$ yields
\[
A_0=e^YY',\qquad B_0=e^{2Y}X'.
\]
These functions are first-order absolute invariants of the limit realization
whose generators are given below. They are functionally independent, since
\[
\det\frac{\partial(A_0,B_0)}{\partial(Y',X')}
=e^{3Y}\ne 0.
\]
In the new coordinates, the invariant differentiation operator $\delta=xD_t$
has the form $\delta_\varepsilon=e^YD_T$; in particular, for the limit
realization we have $\delta_0=e^YD_T$.

In this example, by a limit transition, or contraction, of differential
invariants we mean the limit of a suitably normalized family of invariants in
common coordinates on the jet space, together with verification of the functional
independence of its limit elements. Without additional structure, such a
transition does not define a canonical contraction of an abstract algebra of
invariants.

In the variables $T$, $X$, $Y$, system~\eqref{eq:sys_R1_explicit} for
$\varepsilon\ne 0$ takes the form
\begin{equation}\label{eq:sys_R1_param}
\begin{cases}
Y''=-Y'^2+2\varepsilon\bigl(XY'+X'\bigr)
-2e^{-2Y}\,F({I_\varepsilon},
              {J_\varepsilon}),\\[6pt]
X''=-2\varepsilon X\bigl(XY'+X'\bigr)-2Y'X'
-\dfrac{e^{-3Y}}{\varepsilon}\,
G({I_\varepsilon},{J_\varepsilon}).
\end{cases}
\end{equation}
Notice that in the parameterized system, the arbitrary function $F$ enters only
the equation for $Y''$, while the function $G$ enters only the equation for $X''$.

In the second equation, the contraction parameter $\varepsilon$ appears in the
denominator, so the existence of the limit must be checked for the chosen
function $G$. Suppose $F(I,J)$ and $G(I,J)$ are fixed (i.e., not explicitly
dependent on $\varepsilon$) holomorphic functions near the limit values of their
arguments. A sufficient local condition for regularity is
\[
G(I,0)=0.
\]
Then locally $G(I,J)=JH(I,J)$ for some holomorphic function $H$, and therefore
\[
\frac{G(I_\varepsilon,J_\varepsilon)}{\varepsilon}
=
-e^{2Y}\bigl(X'+\varepsilon X^2\bigr)
H(I_\varepsilon,J_\varepsilon)
\]
has a finite limit. In particular, this condition is satisfied for the fixed
function $G(I,J)=J$.

One may also consider separately a family of functions $G_\varepsilon(I,J)$
depending explicitly on the parameter. For example, if
$G_\varepsilon=\varepsilon\widehat G_\varepsilon$ and the expression
$\widehat G_\varepsilon(I_\varepsilon,J_\varepsilon)$ has a regular limit, then
the corresponding term in the system also has a regular limit. This is a different
formulation from choosing a fixed function $G$.

Choosing the fixed functions $F(I,J)=I$ and $G(I,J)=J$, i.e.,
$F(I_\varepsilon,J_\varepsilon)=I_\varepsilon$ and
$G(I_\varepsilon,J_\varepsilon)=J_\varepsilon$, upon taking the limit
$\varepsilon\to 0$ in system~\eqref{eq:sys_R1_param} we obtain the system
\begin{equation}
\begin{cases}
Y''=-Y'^2+e^{-Y}Y',\\[6pt]
X''=-2 X'Y'+{e^{-Y}}X'.
\end{cases}
\end{equation}

This is just one of the possible choices of functions and the corresponding
limit system. The limit generators of the associated family of realizations
have the form
\[
Q_1=\partial_T,\qquad
Q_2=\partial_X,\qquad
Q_3=T\partial_T-X\partial_X+\partial_Y.
\]
They form the realization $\mathfrak R_1$ of the algebra
$\mathfrak g_{3.4}^{-1}$ from the list~\eqref{real_34}.
Thus, in this example, a transition to $\mathfrak g_{3.4}^{-1}$ rather than to
$\mathfrak g_{3.1}$ has been carried out.

The invariance of the limit system can be verified directly. Denote its
left-hand sides by
\[
L_1=Y''+Y'^2-e^{-Y}Y',\qquad
L_2=X''+2X'Y'-e^{-Y}X'.
\]
Then
\[
Q_1^{(2)}L_a=Q_2^{(2)}L_a=0,\qquad a=1,2,
\]
\[
Q_3^{(2)}L_1=-2L_1,\qquad
Q_3^{(2)}L_2=-3L_2.
\]
Hence, the equation manifold $L_1=L_2=0$ is invariant under the listed generators.

In the example under consideration, the parameterized generators in the
coordinates $(T,X,Y)$ are given by
\[
Q_1^\varepsilon=\partial_T,\qquad
Q_2^\varepsilon=
\varepsilon T^2\partial_T
+(1-2\varepsilon TX)\partial_X
+2\varepsilon T\partial_Y,\qquad
Q_3^\varepsilon=T\partial_T-X\partial_X+\partial_Y.
\]
Their coefficients and the coefficients of their prolongations up to second order
converge regularly to the corresponding coefficients of $Q_1,Q_2,Q_3$. For
$F(I,J)=I$ and $G(I,J)=J$, the right-hand sides of the parameterized system also
have a regular limit together with the derivatives required to verify the
infinitesimal criterion. Thus, the limit transition conditions for this family
are verified, and the direct check above further confirms the invariance of the
limit system.

\section{Concluding remarks}\label{Sec_conclusion}

In this paper, we have investigated limit transitions between local realizations
of complex Lie algebras and invariant systems of ordinary differential equations.
The starting ingredients of the approach are the known chain of contractions of
three-dimensional Lie algebras~\cite{NP2006}, the correspondence between local
transitive realizations and stationary subalgebras, the criterion
$\ker R=\operatorname{core}_{\mathfrak g}(\mathfrak h)$ for a transitive
realization, and the construction of regular intransitive realizations by
promoting parameters of a family of transitive realizations to additional
coordinates. The definition of differential invariants and the criterion
$r_1=r_2$ for a regular normal invariant system were presented
in~\cite{Haponova2006}.

Using the algebra $\mathfrak{sl}(2,\mathbb{C})$ as an example, we constructed
the chain of In\"on\"u--Wigner contractions
\begin{equation*}
\mathfrak{sl}(2,\mathbb{C})\;\longrightarrow\;
\mathfrak{g}^{-1}_{3.4}\;\longrightarrow\;
\mathfrak{g}_{3.1},
\end{equation*}
obtained a list of local realizations and their limit forms, and showed that the
proposed modification of Shirokov's method---consisting in formally lifting
invariant parameters---enables the systematic construction of intransitive
realizations. The entire algorithm reduces to matrix operations---exponentiation,
inversion, and taking limits---without the need to solve differential equations.

Tables~\ref{tab:sl2_to_g34_reals}, \ref{tab:g34_to_g31_reals},
and~\ref{tab:sl2_to_g31_reals} juxtapose the chosen subalgebras, parameterized
vector fields, and their limit expressions. In the relevant cases, local
changes of variables and bases are also provided for comparison with canonical
realizations. The statement regarding faithfulness of a limit transitive
realization applies provided the corresponding limit subalgebra exists and is
compatible with the limit action.

Table~\ref{tab:sl2_to_g34_reals} exhibits an example of \emph{splitting}: two
complex realizations that are locally equivalent for $\varepsilon\ne 0$ yield
inequivalent faithful limit realizations of the algebra $\mathfrak g_{3.4}^{-1}$.
For these limit realizations, the ranks of the action of the derived ideal differ,
being $2$ and $1$, respectively. An example of class merging was not constructed
separately in this work, and therefore merging is not claimed as a demonstrated
result here.

For one generating realization of $\mathfrak{sl}(2,\mathbb C)$ on the regular
domain $x\ne 0$, complete local sets of functionally independent first- and
second-order absolute differential invariants and an invariant differentiation
operator were computed. Using them, a second-order invariant system was written
in normal form and its parameterized family was constructed.

In the example considered, the direct limit of one of the first-order invariants
vanishes; after normalization, two functionally independent invariants of the
limit action are obtained. For the choice $F(I,J)=I$, $G(I,J)=J$, a limit
transition of the systems to a realization of the algebra
$\mathfrak g_{3.4}^{-1}$ was carried out. The existence of the limit depends on
the behavior of the arbitrary functions and does not follow solely from the
contraction of the realizations.

Future research should aim at completing the chains of transitions, classifying
limit stabilizers, analyzing changes in faithfulness and rank, and establishing
the functional completeness of contracted differential invariants.
Particular attention is warranted for the study of singular strata, integrability
and globalization conditions for local actions, compatibility of contractions
with reduction, conditional and hidden symmetries, and the development of
equivalence criteria for classes of systems.

Potential applications include constructing coupled dynamical models with two
degrees of freedom (in particular, Ermakov systems, generalized oscillators, and
central-force problems) and transferring exact or asymptotic solutions across
controlled limits.
Furthermore, the results obtained can serve as a foundation for symbolic
automation of computations, searching for contraction-compatible reductions of
partial differential equations, and analyzing limit regimes in Galilean and
relativistic models.

\subsection*{Acknowledgments}

The authors are grateful to Doctor of Physical and Mathematical Sciences
V.\,M.~Boyko and Candidate of Physical and Mathematical Sciences
B.\,H.~Feshchenko for helpful ideas and remarks during the preparation of this
article.
The authors also thank the anonymous referees for their careful and detailed
analysis of the paper and for comments that helped improve it.

This work was supported by the Simons Foundation
(SFI-PD-Ukraine-00014586, M.S., M.N.); M.N. also gratefully acknowledges financial
support from the National Research Foundation of Ukraine (project no. 2025.07/0405).

\renewcommand{\refname}{References}

\end{document}